\documentclass[slac_one]{revtex4}
\usepackage{graphicx}
\usepackage{fancyhdr}
\usepackage{epsfig}
\renewcommand{\baselinestretch}{1.2}
\def\photonatomrightt{\begin{picture}(3,1.5)(0,0)
                               \put(0,-0.75){\tencirc \symbol{2}}
                               \put(1.5,-0.75){\tencirc \symbol{1}}
                               \put(1.5,0.75){\tencirc \symbol{3}}
                               \put(3,0.75){\tencirc \symbol{0}}
                     \end{picture}
                    }

\def\photonrightthalf{\begin{picture}(15,1.5)(0,0)
                    \multiput(0,0)(3,0){5}{\photonatomrightt}
                 \end{picture}
                }

\def\fermionul{\begin{picture}(15,15)(0,0)
                        \put(0,0){\vector(-1,1){7.5}}
                        \put(-7.5,7.5){\line(-1,1){7.5}}
                  \end{picture}
                 }

\def\fermionur{\begin{picture}(15,15)(0,0)
                        \put(-15,-15){\vector(1,1){7.5}}
                        \put(-7.5,-7.5){\line(1,1){7.5}}
                  \end{picture}
                 }

\def\gaugebosonrighthalf{\begin{picture}(15,1)(0,0)
                            \put(0,0){\line(1,0){0.75}}
                            \multiput(2.25,0)(3,0){4}{\line(1,0){1.5}}
                            \put(14.25,0){\line(1,0){0.75}}
                         \end{picture}
                        }

\def\gaugebosonurhalf{\begin{picture}(15,15)(0,0)
                            \put(0,0){\line(1,1){15.0}}
                  \end{picture}
                 }

\def\gaugebosondrhalf{\begin{picture}(15,15)(0,0)
                            \put(0,0){\line(1,-1){15}}
                  \end{picture}
                 }

\newenvironment{Feynman}[3]{\begin{center}
                            \setlength{\unitlength}{#3 mm}
                            \begin{picture}(#1)(#2)
                            \thicklines
                           }{\end{picture} \end{center}}

\def\cost{\mbox{$\cos\theta_{\rm \tilde t}$}}
\def\fb{\mathrm{\,fb}}

\newcommand{\stp}{\tilde t_1}
\def\mt{\mbox{$m_{\rm \tilde{\rm t}_1}$}}
\def\mchio{\mbox{$m_{\chio}$}}

\def\mchio{\mbox{$m_{\chio}$}}
\def\ifmath#1{\relax\ifmmode #1\else $#1$\fi}%
\def\chio{\ifmath{\mathchoice%
     {\displaystyle\raise.4ex\hbox{$\displaystyle\tilde\chi{^0_1}$}}%
        {\textstyle\raise.4ex\hbox{$\textstyle\tilde\chi{^0_1}$}}%
      {\scriptstyle\raise.3ex\hbox{$\scriptstyle\tilde\chi{^0_1}$}}%
{\scriptscriptstyle\raise.3ex\hbox{$\scriptscriptstyle\tilde\chi{^0_1}$}}}}

\def\mchi{\ifmath{{{m_{\chio}}}}}

\newcommand{\mst}{m_{\rm \tilde{t}_1}}
\newcommand{\neu}{\tilde{\chi}^0}
\newcommand{\mneu}[1]{m_{\tilde{\chi}^0_{#1}}}

\newcommand{\gev}{\,\, \mathrm{GeV}}

\begin{document}
\begin{titlepage}

\thispagestyle{empty}
\def\thefootnote{\fnsymbol{footnote}}       

\begin{center}
\vspace*{-1cm}
\mbox{ }

\end{center}
\begin{flushright}
\Large
\mbox{\hspace{10.2cm} hep-ph/0605225} \\
\end{flushright}
\begin{center}
\vskip 0.2cm
{\Huge\bf
Scalar Top Quark Studies 
}
\vspace{2mm}

{\Huge\bf
with Various Visible Energies
}
\vskip 1cm
{\LARGE\bf A.~Sopczak$^1$, M.~Carena$^2$, A.~Finch$^1$, A.~Freitas$^2$, C.~Milst\'ene$^2$, H.~Nowak$^3$}
\\
\smallskip
\Large 
$\small ^{(1)}$Lancaster University, UK, $\small ^{(2)}$Fermilab, USA, $\small ^{(3)}$DESY Zeuthen, Germany.
\end{center}

\vskip 0.7cm
\centerline{\Large \bf Abstract}

\vskip 6.cm
\hspace*{-0.5cm}
\begin{picture}(0.001,0.001)(0,0)
\put(,0){
\begin{minipage}{\textwidth}
\Large
\renewcommand{\baselinestretch} {1.2}
The precision determination of scalar top quark properties will play an important 
r\^ole at a future International Linear Collider (ILC). Recent and ongoing studies are 
discussed for different experimental topologies in the detector.
First results are presented for small mass differences 
between the scalar top and neutralino masses. This corresponds to a small expected
visible energy in the detector.
An ILC will be a unique accelerator to explore this scenario. 
In addition to finding the existence of light stop quarks, the precise 
measurement of their properties is crucial for testing their impact on the
dark matter relic abundance and the mechanism of electroweak baryogenesis.
Significant sensitivity for mass differences down to 5~GeV are obtained.
The simulation is based on a fast and realistic detector simulation.
A vertex detector concept of the Linear Collider Flavor Identification 
(LCFI) collaboration, which studies pixel detectors for heavy quark flavour identification,
is implemented in the simulations for c-quark tagging.
The study extends simulations for large mass differences (large visible energy)
for which aspects of different detector simulations, 
the vertex detector design, and 
different methods for the determination of the scalar top mass are discussed.
Based on the detailed simulations we study the uncertainties for the dark matter 
density predictions and their estimated uncertainties from various sources. 
In the region of parameters where stop-neutralino co-annihilation leads to a 
value of the relic density consistent with experimental results, as precisely 
determined by the Wilkinson Microwave Anisotropy Probe (WMAP), the stop-neutralino 
mass difference is small and the ILC will be able to explore this region efficiently. 
\renewcommand{\baselinestretch} {1.}

\normalsize
\vspace{0.5cm}
\begin{center}
{\sl \large
Presented at the 13th International Conference on Supersymmetry and Unification of 
                                Fundamental Interactions (SUSY'05), Durham, UK, July 18-23, 2005.
\vspace{-6cm}
}
\end{center}
\end{minipage}
}
\end{picture}
\vfill

\end{titlepage}

\newpage
\thispagestyle{empty}
\mbox{ }
\newpage
\setcounter{page}{0}

\title{{\small{International Conference on Supersymmetry -- SUSY'05, Durham, UK 
}}\\ 
\vspace{12pt}
Scalar Top Quark Studies with Various Visible Energies} 

\author{A.~Sopczak$^1$\footnote{Presented at the 13th International Conference on Supersymmetry and Unification of 
                                Fundamental Interactions (SUSY'05), Durham, UK, July 18-23, 2005.},
        M.~Carena$^2$, A.~Finch$^1$, A.~Freitas$^2$, C.~Milst\'ene$^2$, H.~Nowak$^3$}
\affiliation{(1) Lancaster University, Lancaster LA1 4YB, United Kingdom,}
\affiliation{(2) Fermi National Accelerator Laboratory, Batavia, IL 60510-500, USA,}
\affiliation{(3) Deutsches Elektronen-Synchrotron DESY, D--15738 Zeuthen, Germany.}

\begin{abstract}
The precision determination of scalar top quark properties will play an important 
r\^ole at a future International Linear Collider (ILC). Recent and ongoing studies are 
discussed for different experimental topologies in the detector.
First results are presented for small mass differences 
between the scalar top and neutralino masses. This corresponds to a small expected
visible energy in the detector.
An ILC will be a unique accelerator to explore this scenario. 
In addition to finding the existence of light stop quarks, the precise 
measurement of their properties is crucial for testing their impact on the
dark matter relic abundance and the mechanism of electroweak baryogenesis.
Significant sensitivity for mass differences down to 5 GeV are obtained.
The simulation is based on a fast and realistic detector simulation.
A vertex detector concept of the Linear Collider Flavor Identification 
(LCFI) collaboration, which studies pixel detectors for heavy quark flavour identification, 
is implemented in the simulations for c-quark tagging.
The study extends simulations for large mass differences (large visible energy)
for which aspects of different detector simulations, 
the vertex detector design, and 
different methods for the determination of the scalar top mass are discussed.
Based on the detailed simulations we study the uncertainties for the dark matter 
density predictions and their estimated uncertainties from various sources. 
In the region of parameters where stop-neutralino co-annihilation leads to a 
value of the relic density consistent with experimental results, as precisely 
determined by the Wilkinson Microwave Anisotropy Probe (WMAP), the stop-neutralino 
mass difference is small and the ILC will be able to explore this region efficiently. 
\end{abstract}

\maketitle

\pagestyle{plain}

\section{INTRODUCTION} 

\begin{figure}[tb]
\begin{minipage}{0.49\textwidth}
\begin{center}
\begin{Feynman}{30,0}{45,40}{0.75}
\put(25,40){\fermionul}        \put(4,22){${\rm e^-}$}
\put(25,40){\fermionur}        \put(4,55){${\rm e^+}$}
\put(15,30){\photonrightthalf} \put(32,30){$\gamma$}
\put(25,40){\gaugebosonrighthalf}  \put(30,43){${\rm Z,\gamma^\star}$}
\put(40,40){\gaugebosonurhalf} \put(57,55){${\rm \tilde{t}_1}$}
\put(80,75){\fermionur} \put(83,75){$\chio$ }
\put(80,45){\fermionul} \put(83,45){$\rm c$}
\put(40,40){\gaugebosondrhalf} \put(57,22){$\rm \bar{\tilde{t}}_1$}
\put(80,35){\fermionur} \put(83,35){${\chio}$ }
\put(80,5){\fermionul} \put(83,5){$\rm \bar{c}$}
\end{Feynman}
\end{center}
\end{minipage} \hfill
\begin{minipage}{0.49\textwidth}
\hspace*{-1.3cm}
\includegraphics[scale=0.55]{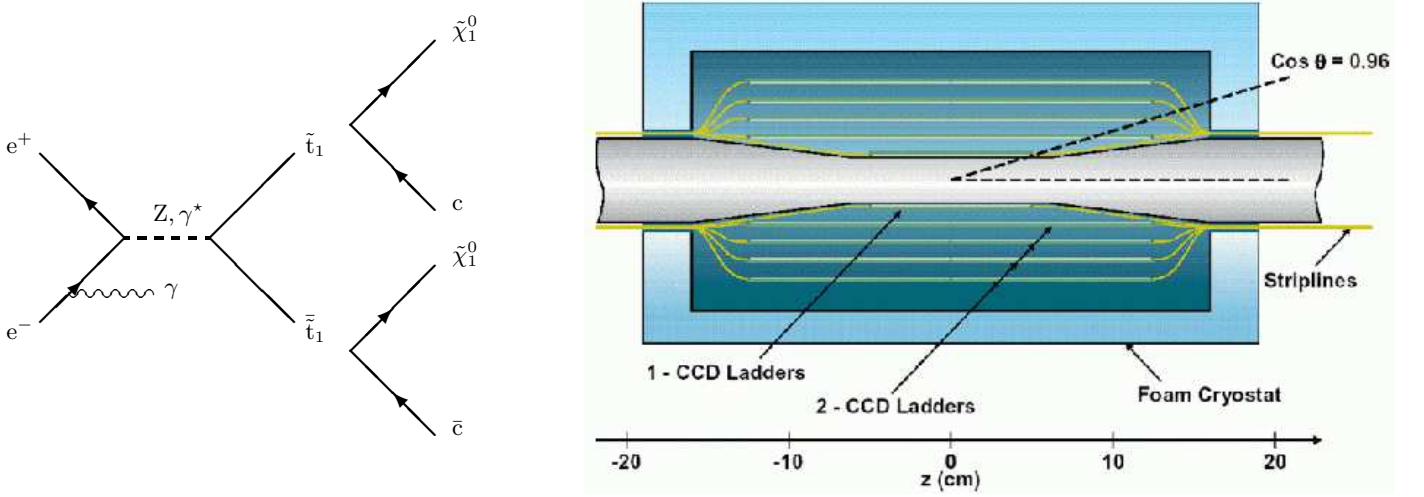}\hfill
\end{minipage}
\caption{
Left: scalar top production and decay.
Right: illustration of the vertex detector with 5 CCD layers.
}
\label{fig:detector}
\end{figure}

The production and decay of scalar top quarks (stops) is particularly interesting
for the development of the vertex detector as only two c-quarks
and missing energy (from undetected neutralinos) are produced for light stops.
The reaction 
$\rm e^+e^- \rightarrow \stp \bar{\tilde{t}}_1 \rightarrow c \chio \bar c \chio$
is shown in Fig.~\ref{fig:detector}. 

The study of small mass differences between stop and neutralino 
is strongly motivated cosmologically.
A long history of experimental observations has corroborated the evidence for
dark matter in the universe, culminating in the recent accurate determination
by the WMAP satellite, in combination with the
Sloan Digital Sky Survey (SDSS)~\cite{Spergel:2003cb}, $\Omega_{\rm CDM} h^2 =
0.1126^{+0.0161}_{-0.0181}$ at the 95\% C.L. Here $\Omega_{\rm CDM}$ is the
dark matter energy density normalized to the critical density and $h$ is the
Hubble parameter in units of 100 km/s/Mpc. Supersymmetry with $R$-parity
conservation provides a natural dark matter candidate, which in most scenarios
is the lightest neutralino.

Electroweak baryogenesis is based on the concept that the
baryon asym\-metry  is generated at the electroweak phase
transition. While in the Standard Model the phase transition is not
sufficiently strongly first order and there is not enough CP violation,
Supersymmetry can alleviate both shortcomings. A strong first-order phase
transition can be induced by loop effects of light scalar top quarks (stops) to
the Higgs potential. In much of the parameter space of interest for electroweak
baryogenesis, the light stop is only slightly heavier than the lightest
neutralino, thus implying that stop-neutralino co-annihilation is significant.
In the  co-annihilation region, the stop-neutralino mass difference is typically
smaller than 30 GeV \cite{Balazs:2004bu}, making a discovery of the stops at hadron
colliders difficult.

The LCFI Collaboration develops a CCD vertex detector for a future Linear Collider.
This vertex detector concept is implemented in the c-quark tagging simulations. 
The detector consists of 5 CCD layers at 15, 26, 37, 48 and 60~mm. 
Figure~\ref{fig:detector} outlines the detector geometry.

Various visible energies in the detector are possible as determined by the mass 
difference between scalar top and neutralino. A small mass difference $\Delta m$ 
corresponds to a small visible energy. Smaller mass differences are a larger 
challenge for the vertex detector as fewer and less energetic tracks are
available to determine the quark flavor.

The work is presented as follows. First, detector simulations with SGV~\cite{berggren} 
and {\sc Simdet}~\cite{simdet} are compared for $\mt = 180$ GeV and 
$\mchio = 100$ GeV. Then, also for large visible energy,
the SPS-5 benchmark parameter point ($\mt = 220.7$ GeV,~$\mchio = 120.0$ GeV) 
has been studied. Different detector design variations are discussed, and four 
methods to determine the scalar top mass are compared.
The experimental simulations of signal and background for small visible 
energy are presented and the scalar top mass and mixing parameter are 
determined for a scenario of the co-annihilation mechanism for Supersymmetric
dark matter. Systematic and statistical uncertainties at a Linear Collider 
are discussed for a dark matter prediction. The expectation from different 
SUSY parameter combinations are compared with the current Cold Dark Matter (CDM) 
measurement.

\section{SGV AND SIMDET COMPARISON FOR LARGE VISIBLE ENERGY}
Signal and background events have been generated for $\sqrt{s} = 500$~GeV
and passed through the {\sc Simdet} 4.03 detector simulation.
First, the 1000 fb$^{-1}$ simulation is compared to a previous SGV simulation
in regard to signal efficiency and numbers of expected background events for 
$\mt = 180$ GeV and $\mchio = 120$ GeV~\cite{andre_lcws04}:
\begin{center}
\vspace*{-0.3cm}
\begin{tabular}{c|c|c|c}
Channel & {\sc Simdet} generated events & {\sc Simdet} preselection/500 fb$^{-1}$ & previous SGV preselection/500 fb$^{-1}$\\
\hline
$\rm c \chio \bar c \chio$   & 50 k    &48\%      & 47\%  \\ \vspace*{-0.2mm}
$\rm q \bar q $ & 12169 k & 64963    & 46788 \\ \vspace*{-0.2mm}
$\rm t \bar t $ & 620 k   & 32715    & 43759 \\ \vspace*{-0.2mm}
$\rm eeZ $ & 5740 k  & 24864    &  4069 \\ \vspace*{-0.2mm}
$\rm ZZ$       & 560 k   &  3100    &  4027 \\ \vspace*{-0.2mm}
$\rm W e \nu $  & 4859 k  & 252367   & 252189\\ \vspace*{-0.2mm}
$\rm WW $  & 6800 k  &122621    & 115243\\ \vspace*{-0.2mm}
Total background    &         &500631    & 466075
\end{tabular}
\vspace*{0.1cm}
\end{center}

The eeZ process has a lower expected rate in SGV, because of a different 
detector coverage in the forward-backward region.
After additional cuts,
$E_{\rm vis}/\sqrt{s} < 0.52$ and $P_{\rm t}/E_{\rm vis} > 0.05$,
the following numbers of events are obtained:
\begin{center}
\begin{tabular}{c|ccccccc}
Channel   &  $\rm q\bar q$ & WW & We$\nu$ & $\rm t\bar t$ & ZZ & eeZ &total  \\ \hline
Background~  & ~6801 & ~23278 & ~226070&   ~5267 & ~125  & ~2147    &~263691
\end{tabular}
\vspace*{0.2cm}
\end{center}
The total number of background events agrees 
well with the previous 278377 events for the SGV simulation~\cite{finch03}.

\begin{figure}[tp]
\begin{center}
\epsfig{file=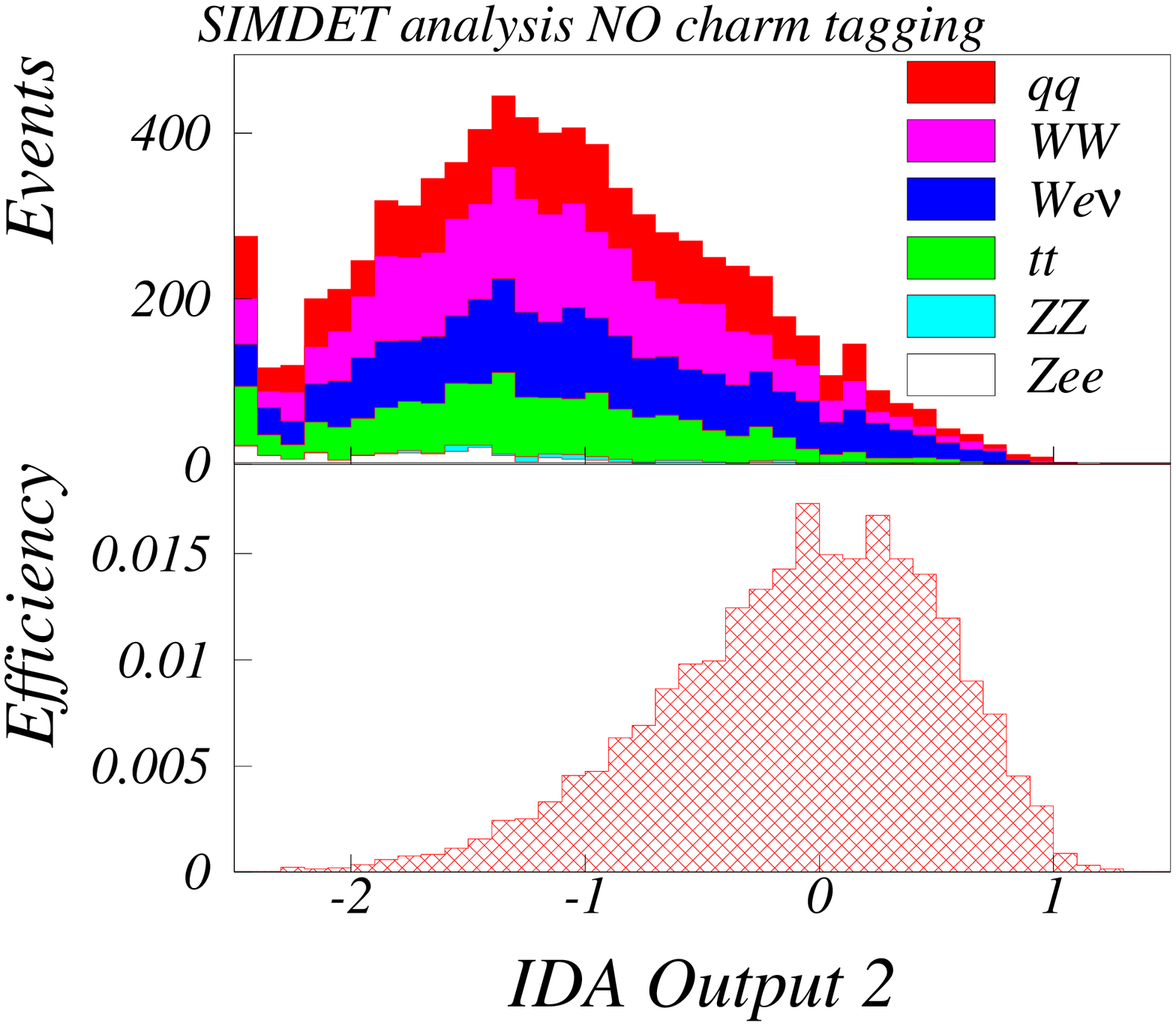, height=5.5cm,width=8cm} \hfill
\epsfig{file=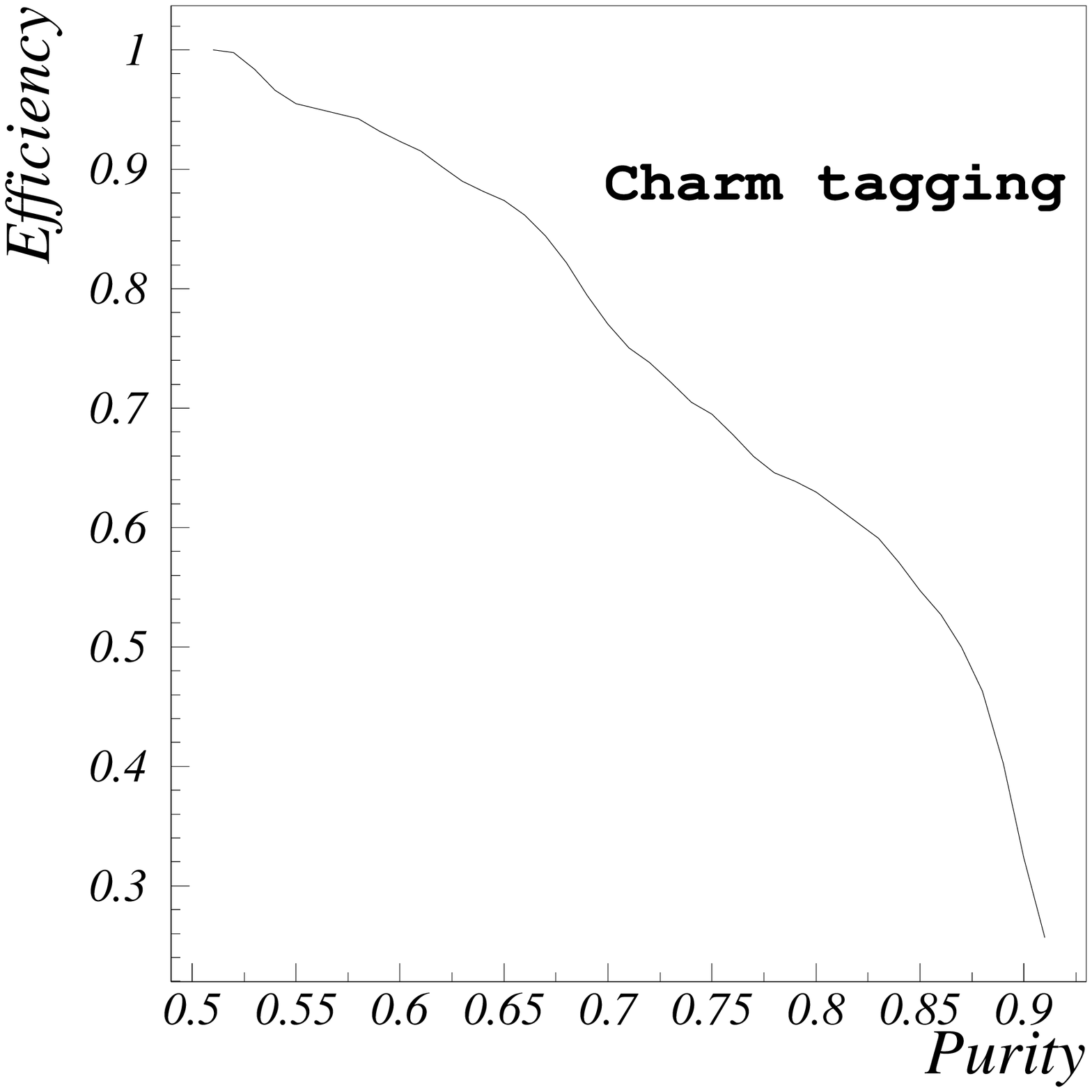,height=5.5cm,width=9cm}
\vspace*{-0.5cm}
\end{center}
\caption{\label{fig:ida}
         Left: IDA-2 output.
         Right: efficiency vs. purity of c-quark tagging after IDA step~1. Purity is defined as 
the ratio of the number of simulated signal events after the selection to all selected events.}
\vspace*{-0.1cm}
\end{figure}

The signal to background ratio is optimized~\cite{finch03} by the IDA method~\cite{IDA}. 
First, by allowing a reduction of the signal of 50\% most background events are removed.
Without c-quark tagging 7815 (cf. SGV 7265) background events remain, 
while with c-quark tagging this number is reduced to 3600 events.
Second, the IDA method is repeated. Figure~\ref{fig:ida} shows the 
background composition after IDA step~2 without c-quark tagging 
and the tagging performance after IDA step~1.
For a 180~GeV signal and 12\% detection efficiency, 680 (cf. SGV 400) background events remain
without c-quark tagging, while with c-quark tagging 165 background events are expected.

\section{SPS-5 (LARGE VISIBLE ENERGY): VERTEX DETECTOR DESIGN VARIATIONS}

The development of a vertex detector for a Linear Collider is large challenge.
A key aspect is the distance of the innermost layer to the interaction point,
which is related to radiation hardness and beam background. Another key aspect
is the material absorption length which determines the multiple scattering. The 
optimization of the vertex detector tagging performance is a further aspect.
While at previous and current accelerators (e.g. SLC, LEP, Tevatron) b-quark 
tagging has revolutionized many searches and measurements, c-quark tagging will 
be very important at a future Linear Collider. Therefore, c-quark tagging could be 
a benchmark for vertex detector developments.

The analysis for a large mass difference with the SPS-5 parameter point (ISAJET)
$\mt = 220.7$~GeV, $\mchi = 120.0$ GeV and $\cost = 0.5377$ was previously 
performed~\cite{andre_lcws04}.
For 25\% (12\%) efficiency 3800 (1800) signal events and 5400 (170) background events 
without c-quark tagging were obtained, 
while the background is reduced to 2300 (68) events with c-quark tagging.

The vertex detector absorption length is varied between normal thickness (TESLA TDR)
and double thickness. In addition, the number of vertex detector layers is varied between
5 layers (innermost layer at 1.5 cm as in the TESLA TDR) and 4 layers (innermost layer at 
2.6 cm).
For SPS-5 parameters the following number of background events remain:
\begin{center}
\vspace*{-3mm}
\begin{tabular}{l|c|cc}
Thickness& layers & 12\% signal efficiency~~~~ &25\% signal efficiency \\
\hline
    Single &  5 (4)  &   68 (82) & 2300 (2681) \\
    Double &  5 (4)  &   69 (92) & 2332 (2765) 
\end{tabular}
\end{center}
As a result, a significantly larger number of background events is expected if the first layer of the 
vertex detector is removed. The distance of the first layer to the interaction point is also
an important aspect from the accelerator physics (beam delivery) perspective. The interplay between
the beam delivery and vertex detector design in regard to critical tolerances like hardware damage of the 
first layer and occupancy (unable to use the data of the first layer) due to beam background goes
beyond the scope of this study and will be addressed in the future.

No significant increase in the expected background is observed by doubling the thickness
of the vertex detector layers. It is interesting to study this behavior for events with smaller visible 
energy in the detector, where a larger effect of the multiple scattering is expected.
This study, based on the analysis in sec.~\ref{sec:low_energy}, is in preparation~\cite{small_deltam_design}.

\section{SPS-5 (LARGE VISIBLE ENERGY): COMPARISION OF MASS DETERMINATIONS}

The precision in the scalar top mass determination at a Linear Collider is crucial and  
four methods are compared for the SPS-5 parameter point~\cite{alex_lcws04}. 
Two of the methods rely on accurate cross section
measurements, the other two use kinematic information from the
observed jets.
The signal events contain two charm jets with large missing energy from the unobserved $\chio$.

\subsection{Mass Determination From Cross Section Measurements}

Accurate cross section measurements combined with theoretical
expectations will make it possible to obtain precise determinations of the scalar top mass.
This requires a high signal sensitivity. An 
Iterative Discriminant Analysis (IDA) method~\cite{IDA} has been used
to obtain a signal to background ratio of 10 or better.
The expected size of the signal is between one thousand and two
thousand events in 500 $\fb^{-1}$ luminosity at a Linear Collider with 
$\sqrt{s} = 500 $ GeV~\cite{finch03}.

\subsubsection{Use of beam polarization }

A high degree of beam polarization is expected to be available at
future $\rm e^+ e^-$ colliders, and by measuring the production cross section in both 
left- and right-handed configurations, $\mt$ and $\cost$ can be determined, as shown in Fig.~\ref{fig:pol} 
and discussed in Ref.~\cite{finch03,bartl97}. For $2 \times 500 \mathrm{\fb}^{-1}$
a precision of $\mt = 220.7 \pm 0.57~\mathrm{GeV}$ and $\cost = 0.538\pm 0.012$ 
is obtained. 

\subsubsection{Threshold scan}
\begin{figure}[ht]
\vspace{-0.75cm}
\epsfig{file=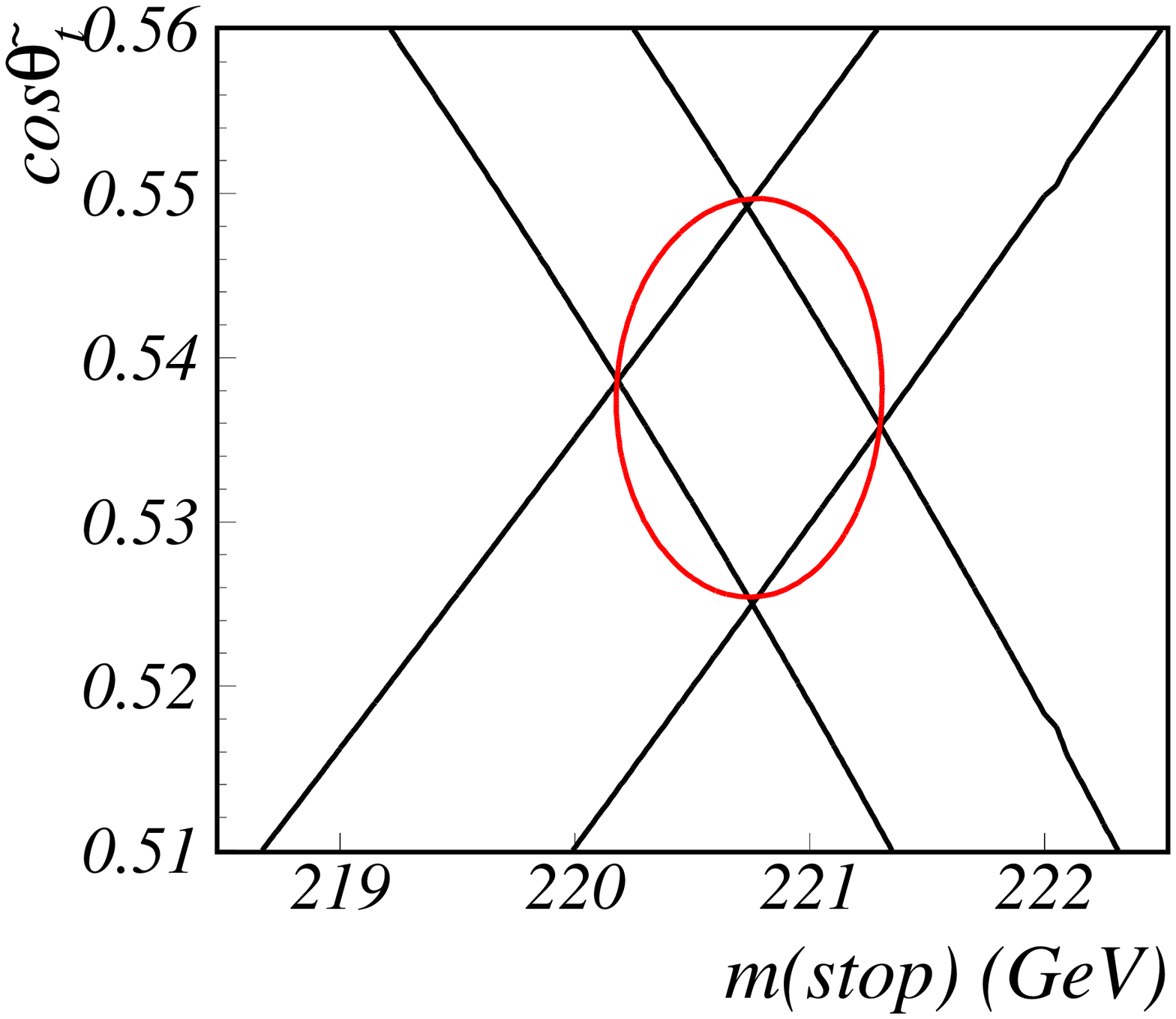,width=8cm,height=5.7cm} \hfill
\epsfig{file=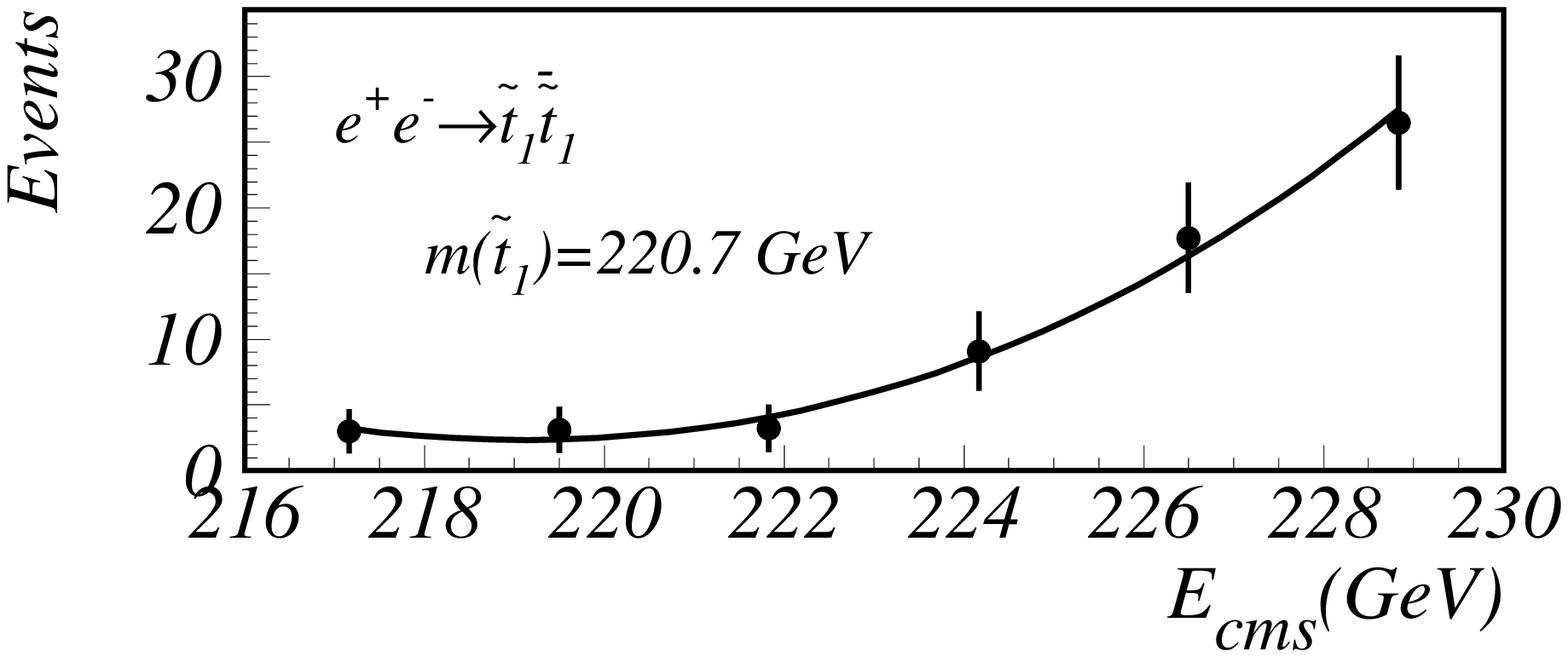,width=9cm,height=5.2cm}
\vspace{-0.6cm}
\caption{Left: $\mt$ and $\cost$ determination from cross section measurements.
         The two bands correspond to different beam polarizations.
         The ellipse indicates the accuracy that could be obtained.
         Right: fit of the scalar top mass from cross section determination 
         near threshold.}
\label{fig:pol}
\end{figure}

Measuring  the cross section for scalar top production close to threshold and 
fitting a theoretical curve allows the mass to be deduced. The excitation curve
near threshold has a $\beta^3$ form. This is shown in Fig.~\ref{fig:pol} for six center-of-mass 
energies, each equivalent to a luminosity of $\mathrm{50\,\fb}^{-1}$. 
In this study the beam polarization $P({\rm e}^-)/P({\rm e}^+)= +80\%/-60\%$
was assumed (right-handed $\rm e^-$) as this provides the best signal to background ratio, leading to 
$\mt = 220.7 \pm 1.2~\mathrm{GeV}$. 

\subsection{Mass Determination From Jet Measurements}

\begin{figure}[ht]
\epsfig{file=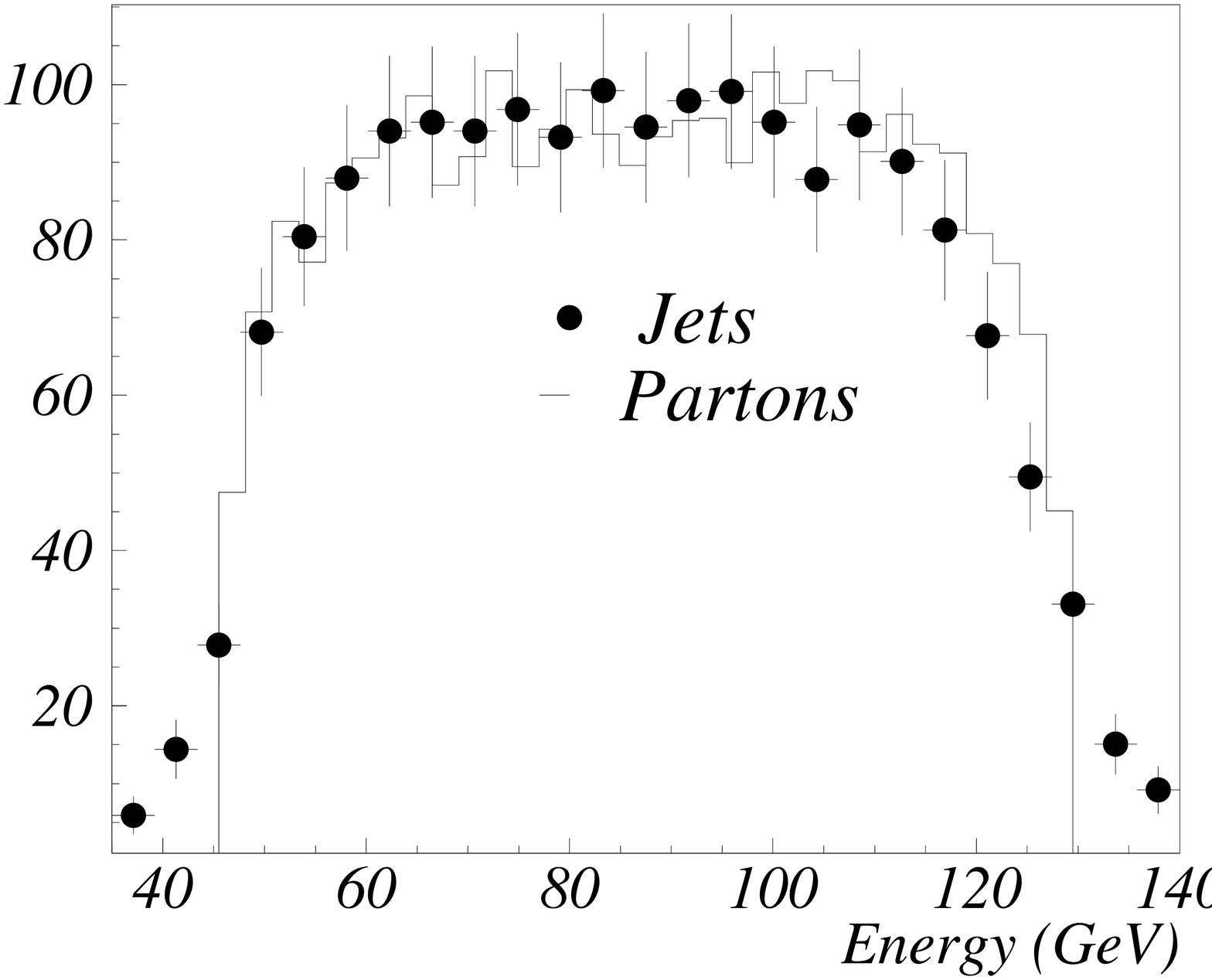,width=0.45\textwidth,height=0.2\textheight} \hfill
\epsfig{file=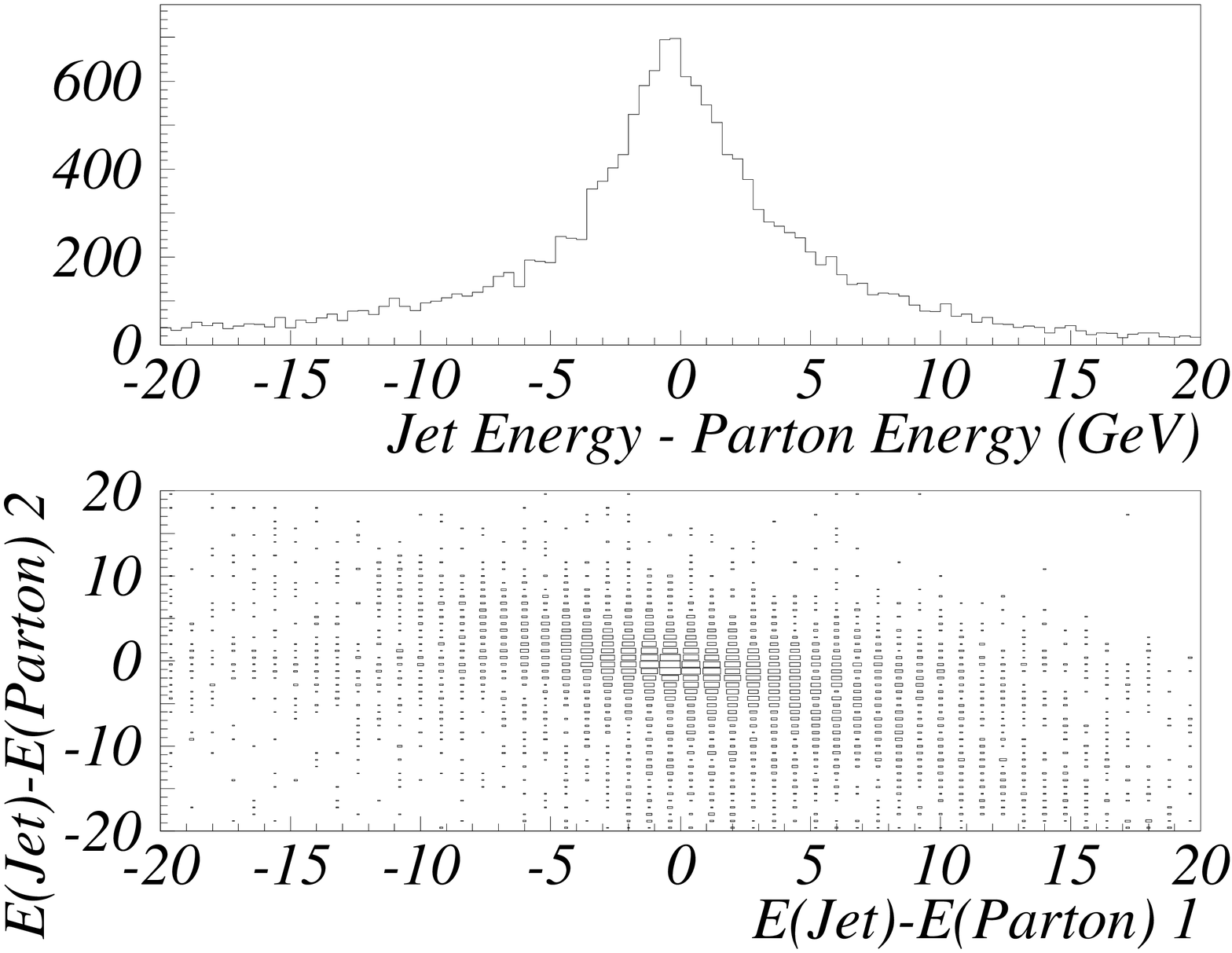,width=0.49\textwidth,height=0.2\textheight}
\vspace{-0.3cm}
\caption{distributions of jet and parton energies.}
\label{fig:endpoint}
\end{figure}

The following two methods rely on measuring the kinematics of the observed jets, 
and thus deriving information about the originating quarks. The precision of this
measurement depends on the jet energy resolution which is expected to be
several GeV in this case as shown in Fig.~\ref{fig:endpoint}.

\begin{figure}[ht]
\begin{minipage}{0.49\textwidth}
\begin{center}
\epsfig{file=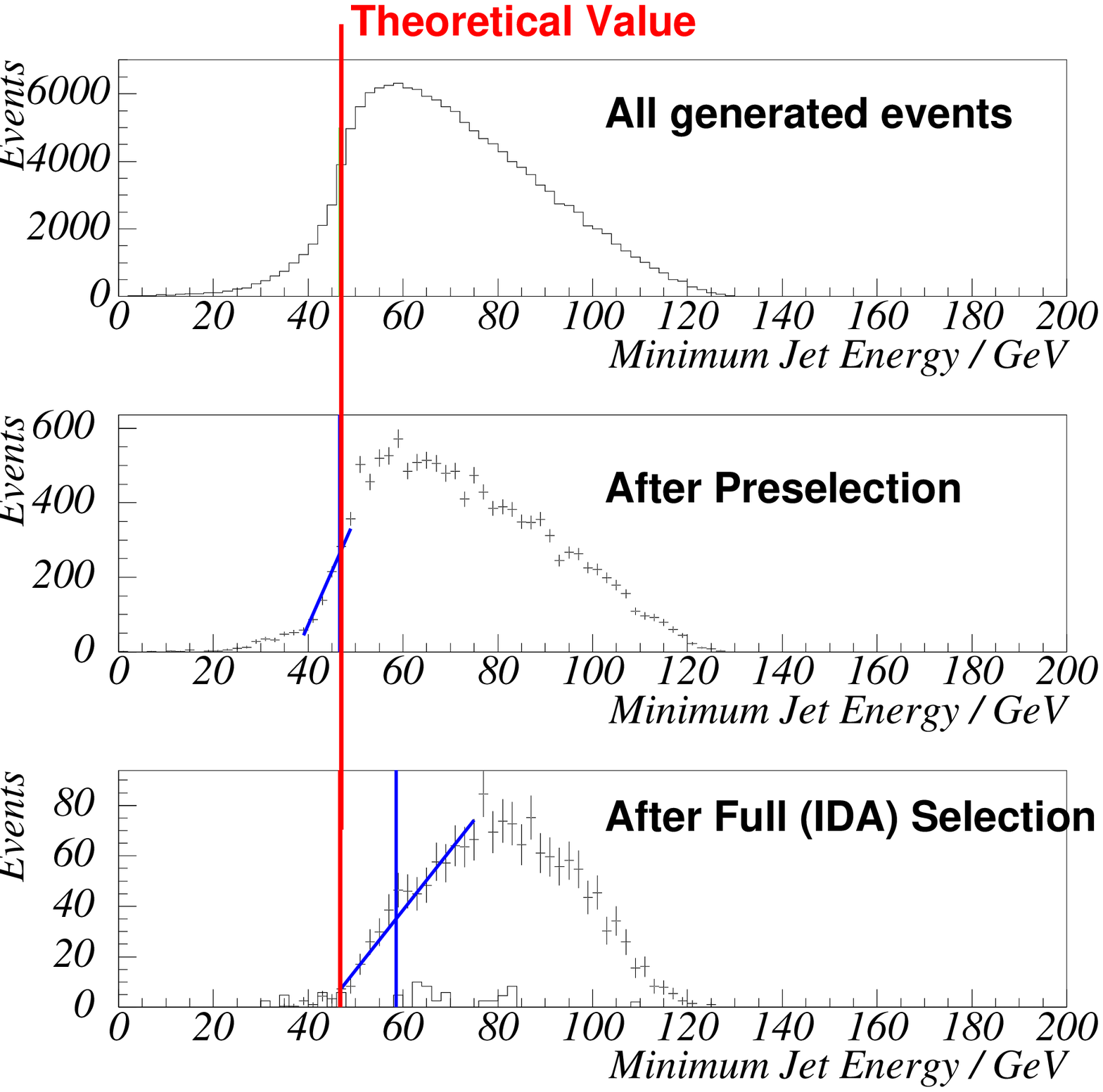,width=\textwidth}
\end{center}
\end{minipage}
\begin{minipage}{0.49\textwidth}
Selection cuts:
\begin{itemize}
\item 20 $<$ number of energy flow objects $<$ 90
\item Visible energy $< 0.8 \sqrt{s}$
\item Longitudinal momentum $< 0.5$ visible energy
\item Thrust $<$ 0.95
\item Cosine of thrust axis relative to beam direction $<$ 0.95
\item Both jet charm tags $>$ 0.3
\item At least one jet charm tag $>$ 0.4
\item Number of jets $<$ 4
\item Lowest energy jet $>$ 35 GeV
\item Highest energy jet $<$  140 GeV
\end{itemize}
\end{minipage}
\vspace*{-3mm}
\caption{Left: distortion of minimum energy spectrum after IDA selection.
         Right: list of sequential selection cuts.}
\label{fig:minjet}
\end{figure}

The IDA method provides an optimal signal to background ratio, but
distorts the jet energy measurements (Fig.~\ref{fig:minjet}), 
thus for this part of the analysis a simple cuts based selection is used.
The cuts are also listed in the figure.
About 900 signal events are selected (11\% efficiency) and 390
background events remain (70\% purity) for unpolarized beams.

\subsubsection{End point method}

\begin{figure}[ht]
\begin{center}
\epsfig{file=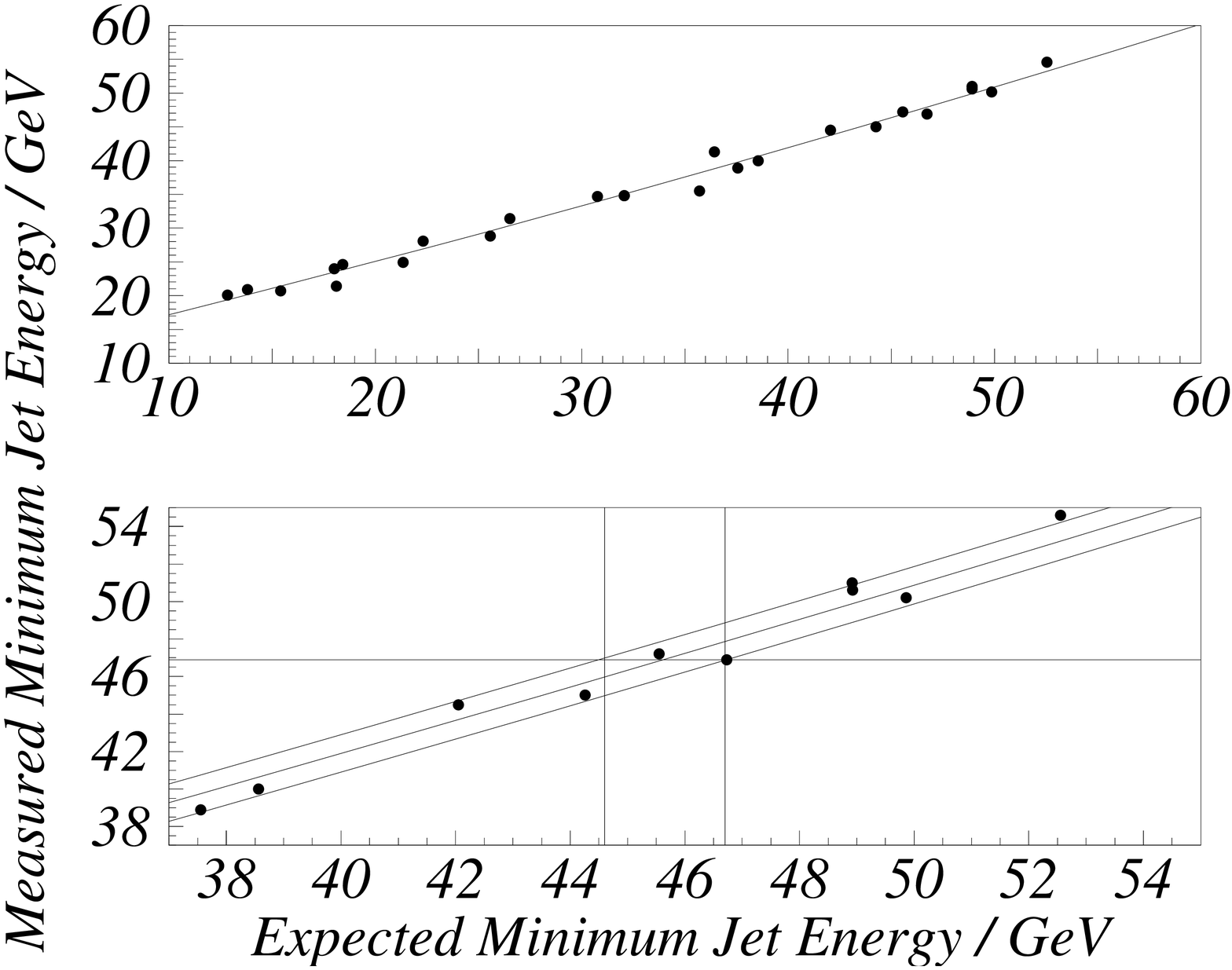,width=0.4\textwidth,height=0.2\textheight}\hfill
\epsfig{file=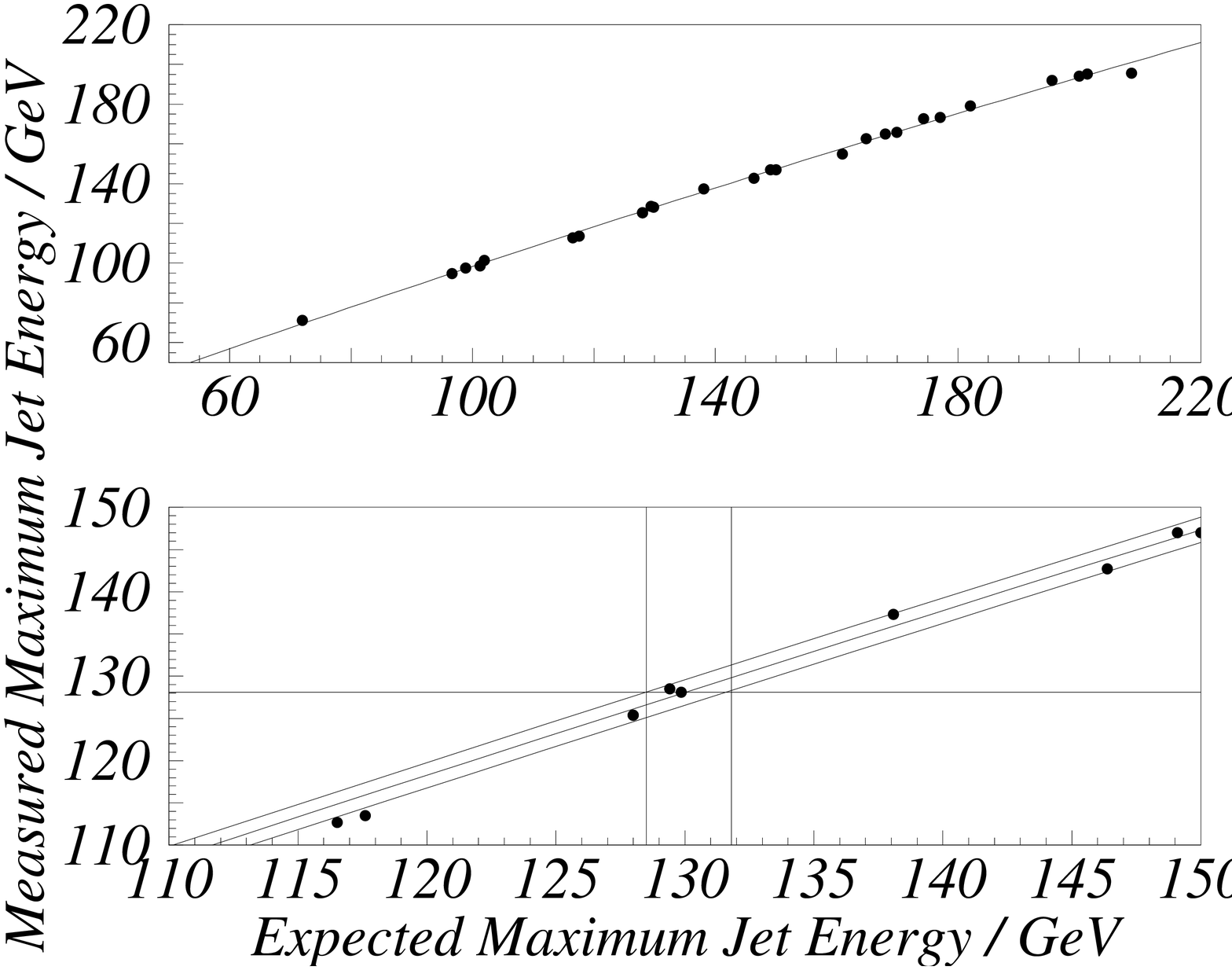,width=0.4\textwidth,height=0.2\textheight}
\vspace{-0.3cm}
\end{center}
\caption{Minimum and maximum jet energy calibration for expected (parton) and simulated (jet) energies.}
\label{fig:jetenergy}
\end{figure}

The energy spectrum of a particle from a two-body decay is approximately a step
function, whose end points contain information about the masses
of both, the particle that decayed, and the other particle produced in
the decay, which in this case is not observed.
In the case of jets, this ideal situation is distorted by detector resolution, 
hadronization, and jet finding, as shown in Fig.~\ref{fig:jetenergy}.
Several event samples are generated to obtain calibration curves and to determine
the mass uncertainty.

The precision for the minimum jet endpoint is $ 45.7 \pm 1.0 $ GeV,
for the maximum jet endpoint $ 130.2 \pm  1.5 $ GeV, and the resulting masses are
$\mt = 219.3 \pm 1.7 $ GeV and $\mchi = 119.4 \pm 1.6 $ GeV.

\subsubsection{Minimum mass method}

\begin{figure}[ht]
\vspace{-0.2cm}
\begin{center}
\epsfig{file=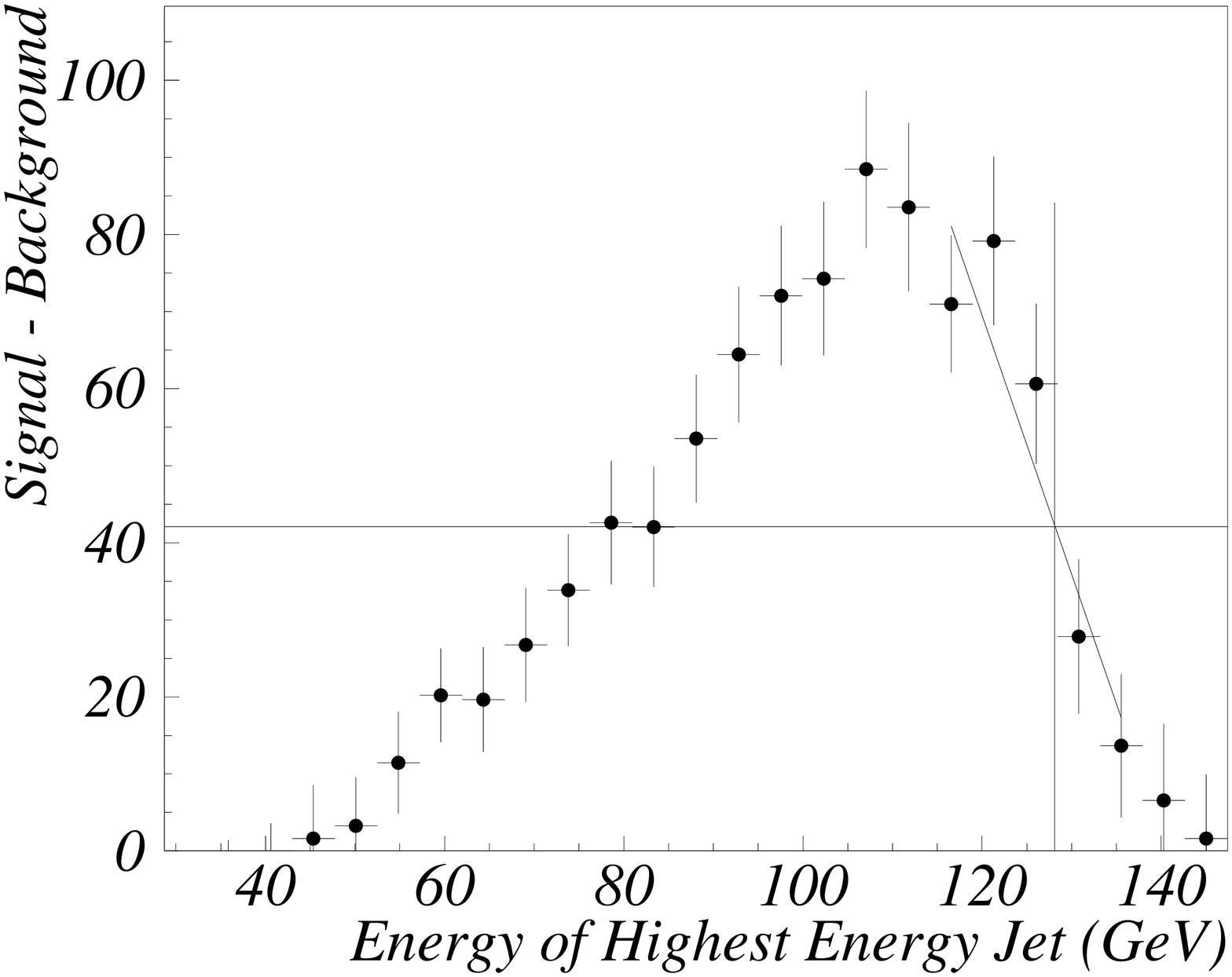,width=7cm,height=4.7cm} \hfill
\epsfig{file=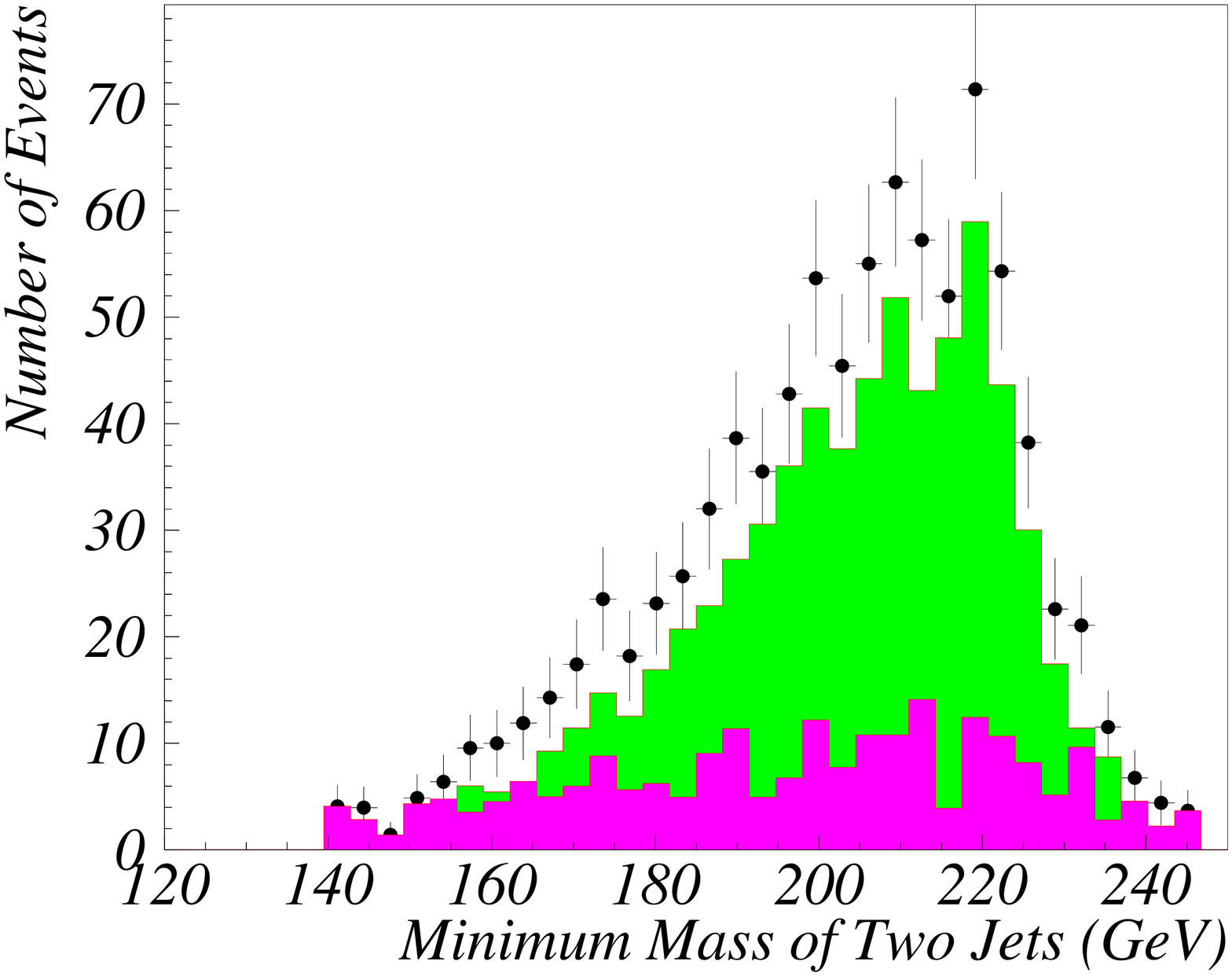,width=7cm,height=4.7cm}
\vspace{-0.6cm}
\end{center}
\caption{Examples of simulated maximum jet energy end points and 
the minimum mass of two jets. 
The points with error bars are the simulated signal. The light gray (green) histogram 
is the scalar top signal and the dark gray (magenta) histogram is the expected background.}
\label{fig:minmass1}
\end{figure}

\begin{figure}[ht]
\begin{center}
\epsfig{file=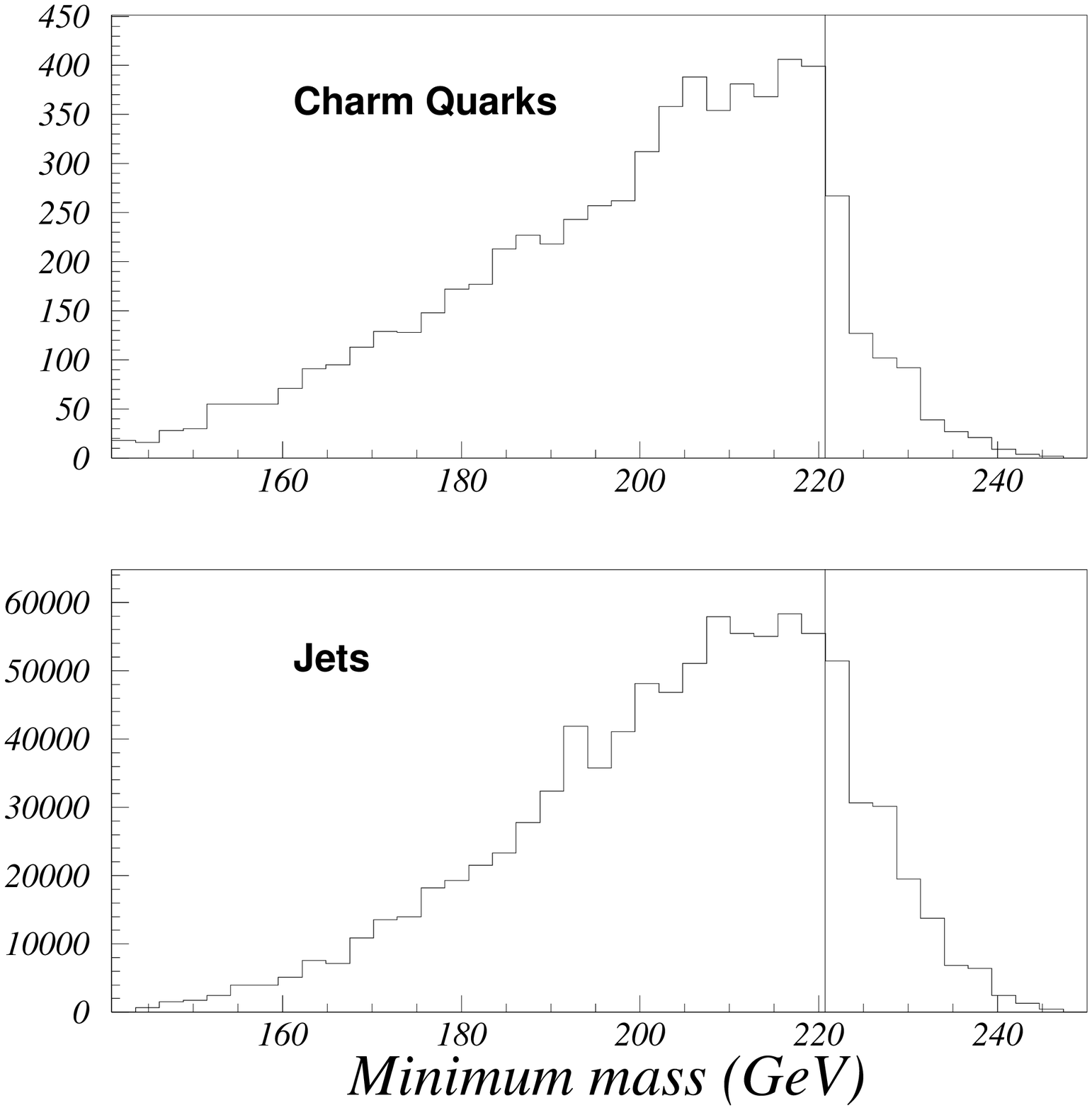,width=0.49\textwidth,height=6cm}
\hfill
\epsfig{file=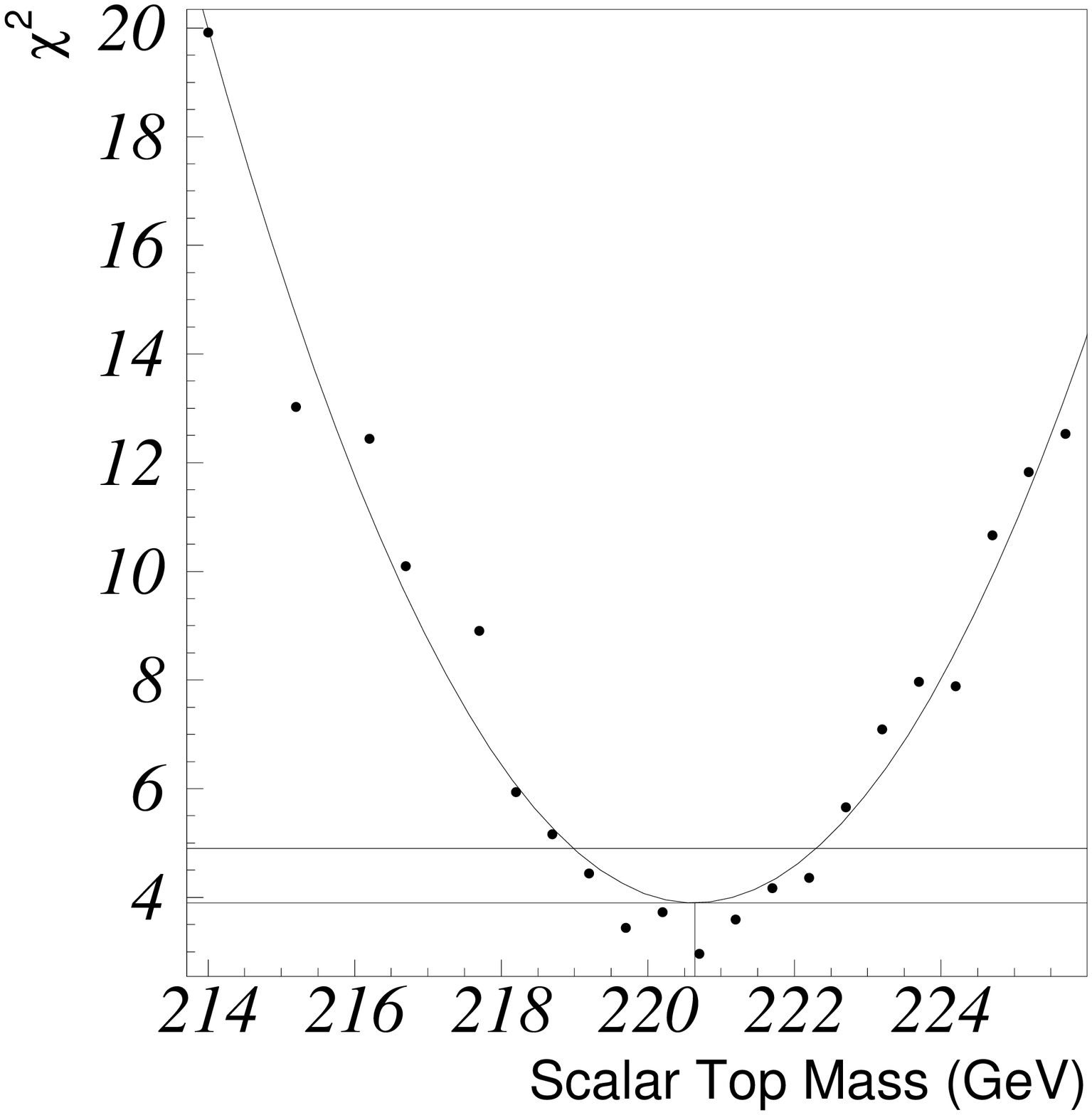,width=0.49\textwidth,height=6cm}
\vspace{-0.6cm}
\end{center}
\caption{Left: minimum mass distribution for partons and jets. 
         Right: $\chi^2$ fit for different scalar top mass simulations.}
\label{fig:minmass2}
\end{figure}

In the case when $\mchi$ is known, the minimum allowed mass of the two jets in 
an event can be calculated~\cite{feng94}. 
Figure~\ref{fig:minmass1} shows an example of the distribution of this variable.
Fitting this distribution with a prediction from simulations allows a precise 
determination of $\mt$, as shown in Fig.~\ref{fig:minmass2}.
This method gives a precision of $\mt = 220.5 \pm 1.5~\mathrm{GeV}$.

\subsection{Discussion of mass determinations}

The precision of the mass determination of the four methods are 
summarized in Table~\ref{tab:massmethods}. 
Slightly higher precision is obtained from the polarization method, however, 
large theoretical uncertainties on the cross section calculations are not included. 
Overall a high sensitivity on the mass determination can be achieved.

\begin{table}[htb]
\caption{Comparison of precision for scalar top mass determination} 
\begin{tabular}{l|c|c|c}
Method & $ \Delta_{\textstyle m}$ (GeV)& luminosity & comment\\
\hline
Polarization & 0.57 & $2 \times 500 \mathrm{\fb}^{-1}$&no theory errors included\\
Threshold scan & 1.2 & $300 \fb^{-1}$ &right-handed $\rm e^-$ polarization\\
End point & 1.7 & $500 \fb^{-1}$&\\
Minimum mass & 1.5 & $500 \fb^{-1}$ & assumes $\mchi$ known\\
\end{tabular}
\label{tab:massmethods}
\vspace{-0.2cm}
\end{table}

\section{SMALL VISIBLE ENERGY STUDIES}
\label{sec:low_energy}

In this section, the production of light stops at a 500 GeV Linear Collider is
analyzed, using high luminosity ${\cal L} = 500 {\rm \
fb}^{-1}$ and polarization of both beams.
The signature for stop pair production at an $\rm e^+e^-$ collider,
\begin{equation}
\rm e^+e^- \to \tilde{t}_1 \, \bar{\tilde{t}}_1 \to c \, \neu_1 \, \bar{c} \, \neu_1,
\end{equation}
is two charm jets and large missing energy. For small $\Delta m$, the jets are
relatively soft and separation from backgrounds is very challenging.
Backgrounds arising from various Standard Model processes can have
cross sections that are several orders of magnitude larger than the signal, so
that even small jet energy variations effects can be important. Thus, it is
necessary to study also this process with a realistic detector simulation.
Signal and background events are generated with {\sc Pythia 6.129}~\cite{pythia}, 
including a scalar top signal generation~\cite{sopczak-stop-gen}
previously used in Ref.~\cite{finch03}.  The detector simulation is
based on the fast simulation {\sc Simdet}~\cite{simdet}, 
describing a typical ILC detector.

Table~\ref{tab:xsec} lists the cross sections for the signal process and the
relevant backgrounds. They have been computed with code used in
Ref.~\cite{slep} and by {\sc Grace 2.0} \cite{grace}, with cross checks to {\sc
CompHep 4.4} \cite{comphep}. A minimal transverse momentum
cut, $p_{\rm t} > 5$ GeV, is applied for the two-photon background, to avoid
the  infrared divergence.
\begin{table}[tb]
\caption{Cross sections for the stop signal and Standard Model background
processes for $\sqrt{s} = 500$ GeV and two polarization combinations. 
The signal is given for the stop mixing angle $\cos \theta_{\tilde{t}} = 0.5$. 
Negative/positive polarization values refer 
to left-/right-handed polarization, respectively.}
\begin{tabular}{l|rrr}
Process &  \multicolumn{3}{c}{cross section [pb]} \\
\hline
$P(e^-) / P(e^+)$ & 0/0 & $\,-$80\%/+60\% & $\,$+80\%/$-$60\% \\
\hline
$\rm \tilde{t}_1 \bar{\tilde{t}}_1 \quad$ $\mst = 120 \gev$ & 0.115 & 0.153 & 0.187 \\
\phantom{$\tilde{t}_1 \tilde{t}_1^* \quad$} $m_{\tilde{t}_1} = 140$ GeV &
  0.093 & 0.124 & 0.151 \\
\phantom{$\tilde{t}_1 \tilde{t}_1^* \quad$} $m_{\tilde{t}_1} = 180$ GeV &
  0.049 & 0.065 & 0.079 \\
\phantom{$\tilde{t}_1 \tilde{t}_1^* \quad$} $m_{\tilde{t}_1} = 220$ GeV &
  0.015 & 0.021 & 0.026 \\
\hline
$\rm W^+W^-$ & 8.55 & 24.54 & 0.77 \\
$\rm ZZ$	& 0.49 & 1.02 & 0.44 \\
$\rm W e\nu$ & 6.14 & 10.57 & 1.82 \\
$\rm e e Z$  & 7.51 & 8.49 & 6.23 \\
$\rm q \bar{q}$, $q \neq t$ & 13.14 & 25.35 & 14.85 \\
$\rm t \bar{t}$ & 0.55 & 1.13 & 0.50 \\
2-photon, $p_{\rm t} > 5$ GeV & 936\phantom{.00}&936\phantom{.00}&936\phantom{.00} \\
\end{tabular}
\label{tab:xsec}
\end{table}

In the first step of the event selection, the following preselection cuts are applied:
\vspace*{-1mm}
\begin{equation}
\begin{array}{lll}
4 < N_{\mbox{\scriptsize charged tracks}} < 50, &\quad &p_{\rm t} > 5 \gev, \\
|\cos \theta_{\rm Thrust}| < 0.8, & & |p_{\rm long,tot}/p_{\rm tot}| < 0.9, \\
E_{\rm vis} < 0.75 \sqrt{s}, & & m_{\rm inv} < 200 \gev.
\end{array}
\end{equation}
The cut on the number of charged tracks removes most leptonic background 
and part of the $\rm t\bar{t}$ background. By requiring a
minimal transverse momentum $p_{\rm t}$, the two-photon background and
back-to-back processes like $\rm q\bar{q}$ are largely reduced. 
The signal is characterized by large missing energy and transverse momentum
from the two neutralinos, whereas for most backgrounds the missing momentum
occurs from particles lost in the beam pipe. Therefore, cuts on the thrust
angle $\theta_{\rm Thrust}$, the longitudinal momentum $p_{\rm long,tot}$, the
visible energy $E_{\rm vis}$ and the total invariant mass $m_{\rm inv}$ are
effective on all backgrounds.
\begin{table}[tb]
\caption{
Left:
background event numbers and $\rm \tilde{t}_1 \bar{\tilde{t}}_1$
signal efficiencies (in \%) for various $\mst$ and $\Delta m$ (in GeV) 
after preselection
and each of the final selection cuts. In the last column the expected event
number are scaled to a luminosity of 500~fb$^{-1}$.
The cuts are explained in the text.
Right:
signal efficiencies (in \%) for $\rm \tilde{t}_1 \bar{\tilde{t}}_1$ production after
final event selection for different combinations of the stop mass $\mst$ and
mass difference $\Delta m = \mst - \mneu{1}$.
}
\begin{tabular}{l|r|r|rrrrrr|r}
 & & after$\;$\rule{0mm}{0mm} &  & &  &  &  & 	& scaled to \\[-.5ex]
Process  & total & presel. & cut 1 & cut 2 & cut 3 & cut 4 & cut 5 & cut 6
	& 500 fb$^{-1}$ \\
\hline
$\rm W^+W^-$ & 210,000 & 2814 & 827 & 28 & 25 & 14 & 14 & 8 & 145 \\ 
$\rm ZZ$	& 30,000 & 2681 & 1987 & 170 & 154 & 108 & 108 & 35 & 257 \\ 
$\rm W e\nu$ & 210,000 & 53314 & 38616 & 4548 & 3787 & 1763 & 1743 & 345 & 5044 \\ 
$\rm e e Z$  & 210,000 & 51 & 24 & 20 & 11 & 6 & 3 & 2 & 36 \\ 
$\rm q \bar{q}$, $q \neq t$ & 350,000 & 341 & 51 & 32 & 19 & 13 & 10 & 8 & 160 \\ 
$\rm t \bar{t}$ & 180,000 & 2163 & 72 & 40 & 32 & 26 & 26 & 25 & 38 \\
2-photon & $3.2 \times 10^6$ & 1499 & 1155 & 1140 & 144 & 101 & 0 & 0 & $<164$\\
\hline
$\mst = 140:$  &&&&&&&&& \\ 
\rule{0mm}{0mm} $\; \Delta m = 20$ & 50,000 & 68.5 & 48.8 & 42.1 & 33.4 &
	27.9 & 27.3 & 20.9 & 9720 \\
\rule{0mm}{0mm} $\; \Delta m = 40$ & 50,000 & 71.8 & 47.0 & 40.2 & 30.3 &
	24.5 & 24.4 & 10.1 & 4700 \\
\rule{0mm}{0mm} $\; \Delta m = 80$ & 50,000 & 51.8 & 34.0 & 23.6 & 20.1 &
	16.4 & 16.4 & 10.4 & 4840 \\
$\mst = 180:$  &&&&&&&&& \\
\rule{0mm}{0mm} $\; \Delta m = 20$ & 25,000 & 68.0 & 51.4 & 49.4 & 42.4 &
	36.5 & 34.9 & 28.4 & 6960 \\
\rule{0mm}{0mm} $\; \Delta m = 40$ & 25,000 & 72.7 & 50.7 & 42.4 & 35.5 &
	28.5 & 28.4 & 20.1 & 4925 \\
\rule{0mm}{0mm} $\; \Delta m = 80$ & 25,000 & 63.3 & 43.0 & 33.4 & 29.6 &
	23.9 & 23.9 & 15.0 & 3675 \\
$\mst = 220:$  &&&&&&&&& \\
\rule{0mm}{0mm} $\; \Delta m = 20$ & 10,000 & 66.2 & 53.5 & 53.5 & 48.5 &
	42.8 & 39.9 & 34.6 & 2600 \\
\rule{0mm}{0mm} $\; \Delta m = 40$ & 10,000 & 72.5 & 55.3 & 47.0 & 42.9 &
	34.3 & 34.2 & 24.2 & 1815 \\
\rule{0mm}{0mm} $\; \Delta m = 80$ & 10,000 & 73.1 & 51.6 & 42.7 & 37.9 &
	30.3 & 30.3 & 18.8 & 1410 
\end{tabular} \hfill
\begin{tabular}{r|cccc}

$\Delta m$ & \multicolumn{4}{|c}{$m_{\tilde{\rm t}_1}$ (GeV)}  \\
(GeV) & 120 & 140 & 180 & 220 \\
\hline
80 & & 10 & 15 & 19 \\
40 & & 10 & 20 & 24 \\
20 & 17 & 21 & 28 & 35 \\
10 & 19 & 20 & 19 & 35 \\
5  &  2.5 & 1.1 & 0.3 & 0.1 
\end{tabular}

\label{tab:evtn}
\end{table}
The various background are substantially reduced after these preselection cuts, 
while about 70\% of the signal is preserved, as shown in Table~\ref{tab:evtn}.

After generating large event samples with the preselection cuts
for the various backgrounds, as listed in Table~\ref{tab:evtn},
the following final event selection cuts are applied to further improve the
signal-to-background ratio:
\begin{enumerate}

\item Number of jets $N_{\rm jets} = 2$. Jets are
reconstructed with the Durham algorithm with the jet resolution parameter
$y_{\rm cut} = 0.003 \times \sqrt{s}/E_{\rm vis}$. The cut
reduces substantially the number of W and quark-pair events.

\item Large missing energy, $E_{\rm vis} < 0.4 \sqrt{s}$. This cut is
effective against $\rm W^+W^-$, $\rm ZZ$ and di-quark events. 
In addition, a window for the invariant jet mass around the
W-boson mass, 
$70 < m_{\rm jet,inv} < 90 \gev$,
is excluded to reduce the large $\rm W e \nu$ background.

\item The number of $\rm q\bar{q}$ events are reduced by requiring a minimal acollinearity angle  
$\cos \phi_{\rm aco} > -0.9$. 

\item Cutting on the thrust angle, $|\cos \theta_{\rm Thrust} < 0.7|$, reduces
W boson background.

\item A strong cut on the transverse momentum, $p_{\rm t} > 12 \gev$,
completely removes the remaining two-photon events.

\item The largest remaining background is from $\rm e^+e^- \to W e\nu$. It
resembles the signal closely in most distributions, e.g. as a function of
the visible energy, thrust or acollinearity. By increasing the invariant
jet mass window from cut 2 to ($60 < m_{\rm jet,inv} < 90 \gev$), the
signal-to-background ratio is improved, but at the cost of a substantial
signal reduction. In addition, the signal selection is enhanced by
c-quark tagging, which is implemented based on the neural network analysis
described in Ref.~\cite{kuhlhiggs}. The neural network has been optimized to
reduce the $\rm We\nu$ background while preserving the stop signal for small mass
differences. 
\vspace*{-3mm}
\end{enumerate}

The resulting event numbers, scaled to a luminosity of 500~fb$^{-1}$, and the
signal efficiencies are listed in Table~\ref{tab:evtn}. After the final
selection, the $\rm \tilde{t}_1 \bar{\tilde{t}}_1$ signal event numbers are of
the same order as the remaining background, $N \sim {\cal O}(10^4)$.

\begin{figure}[htb]
\epsfig{file=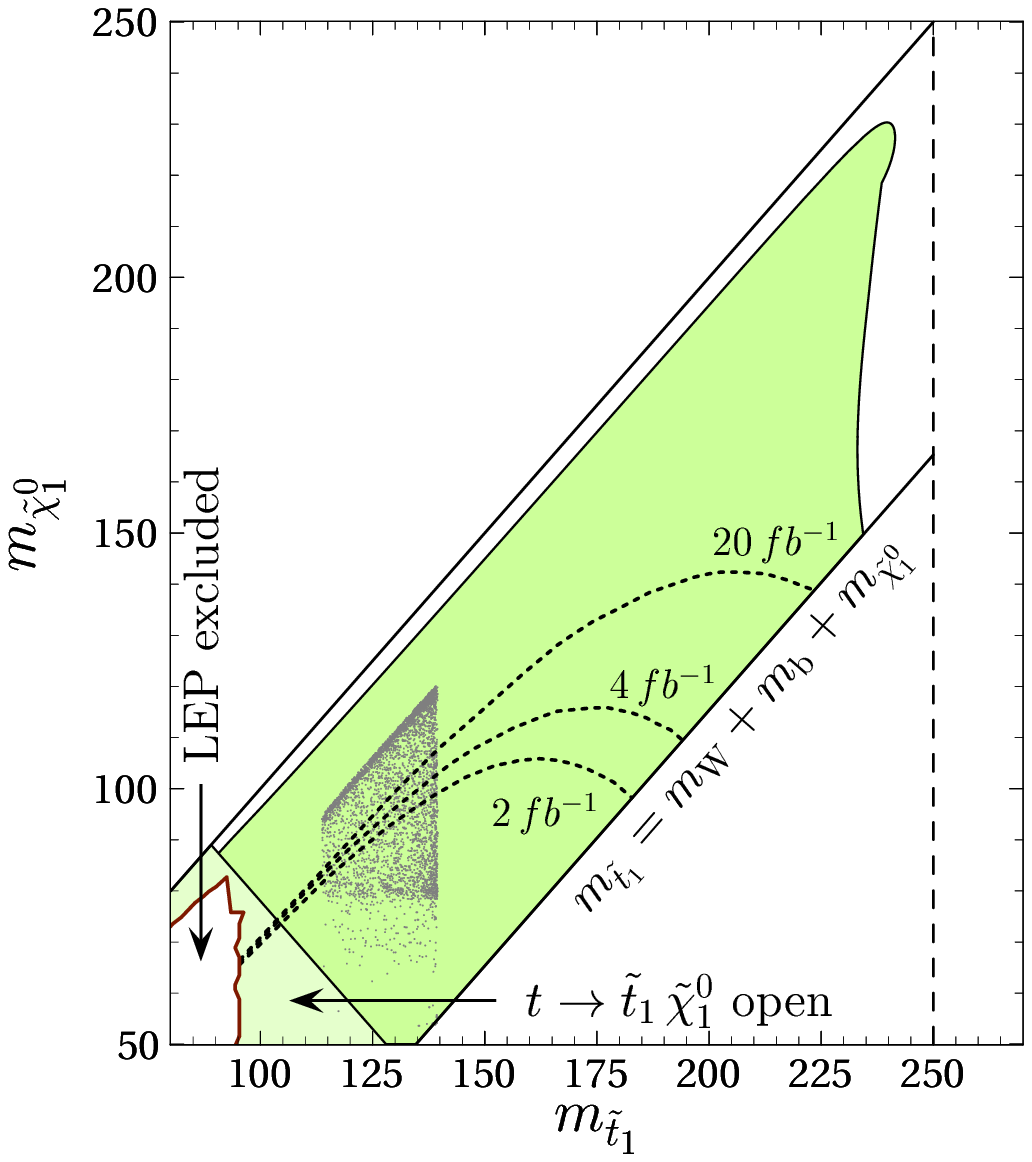, height=8cm, width=8.5cm} \hfill
\epsfig{file=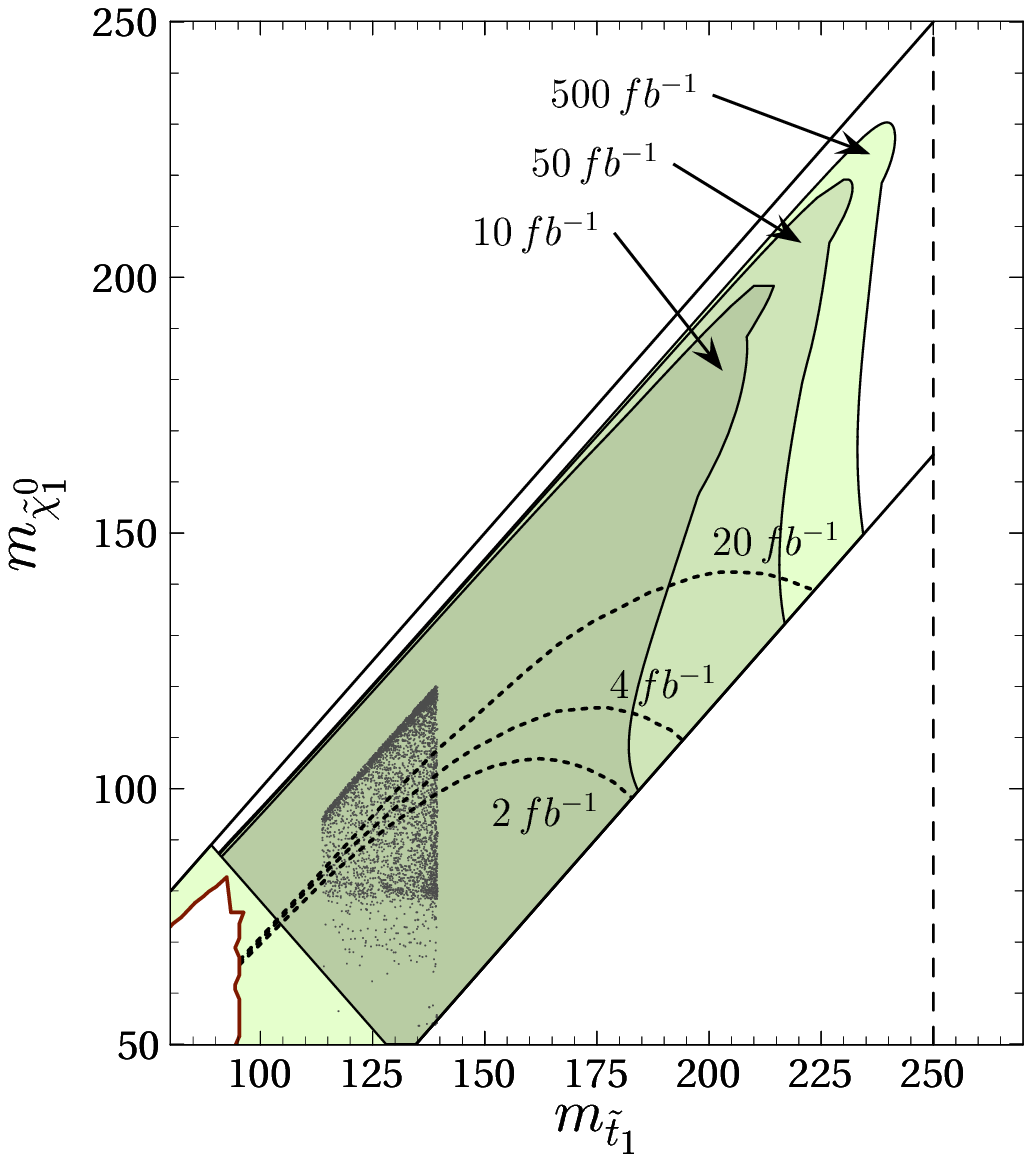, height=8cm, width=8.5cm}
\vspace*{-3mm}
\caption{
Left: discovery reach of a Linear Collider with 500 fb$^{-1}$ luminosity at
$\sqrt{s} = 500$ GeV for the reaction 
$\rm e^+e^- \to \tilde{t}_1 \, \bar{\tilde{t}}_1 \to c \neu_1 \, \bar{c} \,\neu_1$. 
The results
are given in the stop vs. neutralino mass plane. In the gray shaded region, a
5$\sigma$ discovery is possible. The region $\mneu{1} > \mst$ is
inconsistent with a neutralino as Lightest Supersymmetric Particle (LSP), while for 
$\mst > m_{\rm W} + m_{\rm b} + m_{\neu_1}$ 
the three-body decay 
$\rm \tilde{t}_1 \to W^+ \bar{b} \neu_1$ becomes accessible and
dominant.  In the light shaded corner to the lower left, the decay of the top
quark into a light stop and neutralino is open.
The dark gray dots indicate the region consistent with baryogenesis and dark matter.
Also shown are the parameter
region excluded by LEP searches \cite{lep} (white area in the lower left corner) and
the Tevatron light stop reach \cite{Demina:1999ty} (dotted lines) for various
integrated luminosities.
Right: discovery reach for different luminosities.
}
\label{fig:cov}

\end{figure}

To explore the reach for very small mass differences  
$\Delta m = \mst - \mneu{1}$, signal event samples have been generated also for  
$\Delta m = 10 \gev$ and 5 GeV, as shown in Fig.~\ref{fig:cov}, together with results
for larger mass differences. 
The signal efficiency drastically drops for $\Delta m = 5 \gev$ as a result of 
the $p_{\rm t}$ cut (cut 5). An optimization of the event selection for very 
small $\Delta m$ will be addressed in future work.

Based on the above results from the experimental simulations, the discovery 
reach of a 500 GeV $\rm e^+e^-$ collider can be estimated (Fig.~\ref{fig:cov}).
The signal efficiencies for the parameter points in 
Fig.~\ref{fig:cov} 
are interpolated to cover the whole parameter region.
The expected signal rates $S$ are computed
for each mass combination $(\mst,\mneu1)$.
Together with the number of background events $B$,
this yields the significance $S/\sqrt{S+B}$. The gray (green) area in the figure
corresponds to the $5\sigma$ discovery region, $S/\sqrt{S+B} > 5$.

As evident from the figure, the ILC can find light stop quarks for mass differences 
down to $\Delta m \sim {\cal O}(5 \gev)$, beyond the stop-neutralino
co-annihilation region. The figure (right plot) shows the reach
which can be achieved with small total luminosities.

\section{STOP PARAMETER DETERMINATION}

The discovery of light stops would hint towards the possibility of
electroweak baryogenesis and may allow the co-annihilation mechanism 
to be effective.
In order to confirm this idea, the relevant
Supersymmetry parameters need to be measured accurately. 
In this section, the experimental determination of
the stop parameters will be discussed.

For definiteness, a specific MSSM parameter point is chosen:
\begin{equation}
\begin{array}{rlrlrlrl}
m^2_{\rm\tilde{U}_3} &= -99^2 \gev^2,	
& A_t &= -1050 \gev,
& M_1 &= 112.6 \gev, 
& |\mu| &= 320 \gev, \\
m_{\rm\tilde{Q}_3} &= 4200 \gev,	
& \tan\beta &= 5,
& M_2 &= 225 \gev,
& \phi_\mu &= 0.2.
\end{array}
\label{eq:scen}
\end{equation}
The chosen parameters are compatible with the mechanism of electroweak
baryogenesis, generating the baryon asymmetry through the phase of $\mu$. They
correspond to a value for the dark matter relic abundance
within the WMAP bounds, $\Omega_{\rm CDM} h^2 = 0.1122$. The relic
dark matter density has been computed with the code used in Ref.~\cite{morr}.
In this scenario, the
stop and lightest neutralino masses are $m_{\rm \tilde{t}_1} = 122.5 \gev$ and
$\mneu{1} = 107.2 \gev$, and the stop mixing angle is 
$\cos \theta_{\rm \tilde{\rm t}} = 0.0105$, 
i.e. the light stop is almost completely right-chiral. The
mass difference $\Delta m = m_{\rm \tilde{t}_1} - \mneu{1} = 15.2 \gev$ lies within
the sensitivity range of the ILC.

The measurements of $\rm \tilde{t}_1 \bar{\tilde{t}}_1$ production cross sections for
different beam polarizations make it possible to extract both the mass of the light stop
and the stop mixing angle~\cite{bartl97}.  Here is it assumed that 250
fb$^{-1}$ is spent each for $P({\rm e}^-)$/$P({\rm e}^+) = -80\% / +60\%$ and 
$+80\%$/$-60\%$,
where negative/positive polarization degrees indicate left-/right-handed
polarization. In the cross section measurements the statistical and systematic 
errors are similar and of about 0.8\% each.
A complete discussion of the systematic errors is given in Ref.~\cite{paper}.

\begin{figure}[tb]
\begin{minipage}{0.49\textwidth}
\epsfig{file=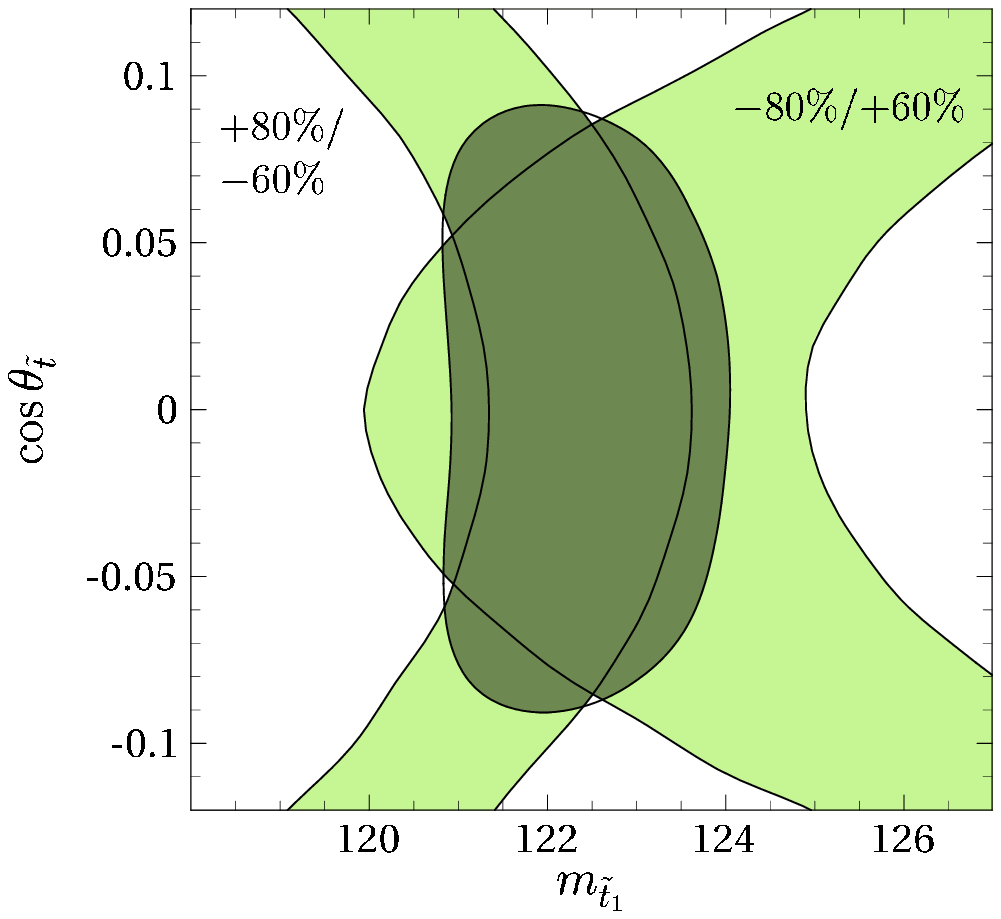, height=8cm, width=\textwidth, bb=20 422 306 682}
\end{minipage}
 \hfill
\begin{minipage}{0.49\textwidth}
\epsfig{file=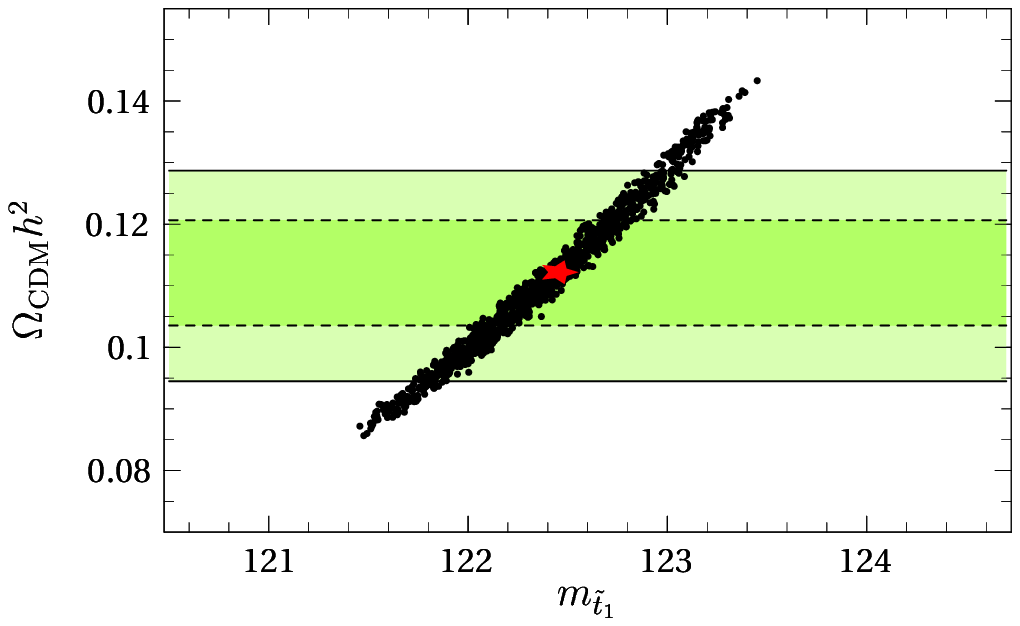, height=8cm, width=\textwidth}
\end{minipage}
\vspace*{0.5cm}
\caption{
Left:
determination of light stop mass $\mst$ and stop mixing angle
$\theta_{\tilde{\rm t}}$ from measurements of the cross section 
$\sigma(\rm e^+e^- \to \tilde{t}_1 \bar{\tilde{t}}_1)$ 
for beam polarizations $P({\rm e^-})/P({\rm e^+}) = -80\%/+60\%$ and
$+80\%/-60\%$. Statistical and systematic errors are included.
Right:
computation of dark matter relic abundance $\Omega_{\rm CDM} h^2$
taking into account estimated experimental errors for stop, chargino, neutralino
sector measurements at future colliders. The black dots correspond to
a scan over the 1$\sigma$ ($\Delta \chi^2 \leq 1$) region which is allowed by the 
expected experimental errors, as a
function of the measured stop mass. The red star indicates the best-fit
point. The horizontal shaded bands show the
1$\sigma$ and 2$\sigma$ constraints on the relic
density measured by WMAP.}
\label{fig:stoppar}
\end{figure}
Each of the two cross section measurements for $P({\rm e}^-)/P({\rm e}^+) = -80\%/+60\%$
and $+80\%/-60\%$ corresponds to a band in the parameter plane of the stop mass and
mixing angle, as shown in Fig.~\ref{fig:stoppar}. Combining the two
cross section measurements, the stop parameter are determined to
\begin{equation}
m_{\rm \tilde{t}_1} = 122.5 \pm 1.0 \gev, \qquad
\cos \theta_{\rm \tilde{t}} < 0.074 \quad \Rightarrow
\sin \theta_{\rm \tilde{t}} > 0.9972.
\end{equation}

\begin{figure}[tb]
\begin{minipage}{0.49\textwidth}
\begin{tabular}{l|cccccc}
 &  A & B & C & D & E & F \\
\hline
$m^2_{\rm\tilde{U}_3}$ [GeV$^2$] & 
  $-99^2$ & $-99^2$ & $-99^2$ & $-97^2$ & $-90.5^2$ & $-85.5^2$ 
\\
$m_{\rm\tilde{Q}_3}$ [GeV] &
  2700 & 3700 & 4200 & 4900 & 4700 & 4300
\\
$A_t$ [GeV] &
  $-860$ & $-1150$ & $-1050$ & $-500$ & $-400$ & 0
\\
$M_1$ [GeV] &
  107.15 & 111.6 & 112.6 & 119.0 & 123.2 & 129.0 
\\
$\tan\beta$ &
  5.2 & 4 & 5 & 6 & 5.5 & 5.5 
\\
$A_{e,\mu,\tau} \times e^{i \pi/2}$ & 
5  & 3.7 & 5 & 5.8 & 5.2 & 5 \\
\hline
$m_{\tilde{t}_1}$ [GeV] &
  117.1 & 118.0 & 122.5 & 130.2 & 135.2 & 139.4
\\
$\mneu{1}$ [GeV] &
  102.1 & 104.1 & 107.2 & 114.0 & 118.1 & 123.1
\\
$\cos \theta_{\tilde{\rm t}}$ &
  0.0210 & 0.0150 & 0.0105 & 0.0038 & 0.0035 & 0.0005
\\
$m_{\rm h^0}$ [GeV] &
  115.1 & 115.0 & 117.0 & 117.1 & 116.2 & 115.1
\\
$\Omega_{\rm CDM} h^2$ &
  0.113 & 0.060 & 0.112 & 0.144 & 0.166 & 0.112 
\end{tabular}
\end{minipage}
\begin{minipage}{0.49\textwidth} \hfill
\includegraphics[height=8cm,width=1\textwidth]{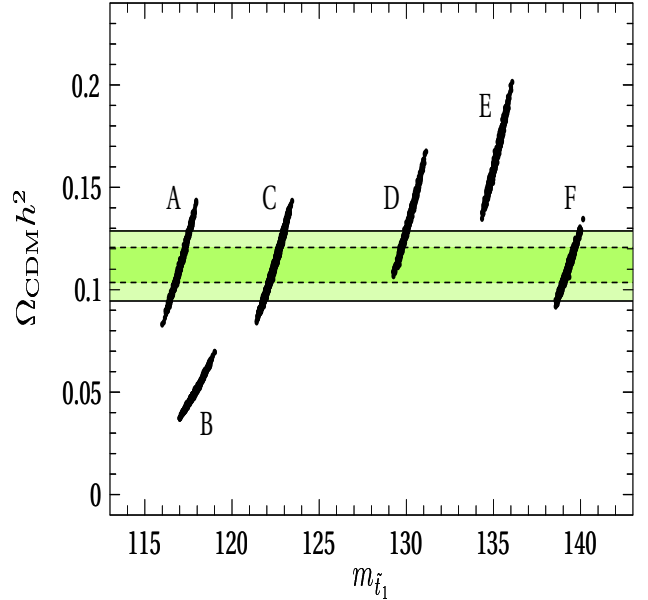}
\end{minipage}
\vspace*{-2mm}
\caption{
Left: dark matter scenarios in the Supersymmetric model.
      Point C corresponds to the scenario discussed before.
Right: dark matter relic abundance $\Omega_{\rm CDM} h^2$ for WMAP 1$\sigma$ and 2$\sigma$ 
error bands and expected Linear Collider precision for benchmarks A-F.}
\label{fig:dm2}
\end{figure}

The mass of the heavier stop $\rm \tilde{t}_2$ is too large to be
measured directly, but it is assumed that a limit of $m_{\rm \tilde{t}_2} > 1000$
GeV can be set from collider searches.
Combining the stop parameter measurements with corresponding data from the neutralino
and chargino sector~\cite{paper} makes it possible to compute the neutralino dark matter abundance
from experimental results in the MSSM.
All experimental errors are propagated and correlations are taken into account
by means of a $\chi^2$ analysis. The result of a scan over 100000 random points
in the parameter space allowed by the expected experimental uncertainties
for the scenario eq.~(\ref{eq:scen}) is shown in Fig.~\ref{fig:stoppar} as a function
of the scalar top quark mass.  
The horizontal bands depict the relic density as measured by WMAP~\cite{Spergel:2003cb}, 
which is at the 1$\sigma$ level $0.104 < \Omega_{\rm CDM} h^2 < 0.121$. 
Further scenarios are investigated and compared to the WMAP measurement (Fig.~\ref{fig:dm2}).

The collider measurements of the stop and chargino/neutralino
parameters constrain the relic density to $0.100 < \Omega_{\rm CDM} h^2 <
0.124$ at the 1$\sigma$ level, with an overall precision comparable
to the direct WMAP determination. 

\section{CONCLUSIONS}

Scalar top quark production and decay at a Linear Collider are studied with 
a realistic detector simulation with focus on the c-tagging performance of a 
CCD vertex detector.
The {\sc Simdet} simulation largely agrees with the previous SGV simulation in
the kinematic distributions. In addition, the {\sc Simdet} simulation includes 
a CCD vertex detector (LCFI Collaboration).
The tagging of c-quarks reduces the background by about a factor 3 in the 
$\rm c\tilde\chi^0_1 \bar c\tilde\chi^0_1$ channel.
Thus, scalar top processes can serve well as a benchmark reaction for the 
vertex detector performance.

Dedicated simulations with SPS-5 parameters are performed.
The expected background depends significantly on the detector design,
mostly on the radius of the inner layer.
Future studies of different detector designs will include simulations with small
scalar top and neutralino mass differences.

For the scalar top mass determination four methods are compared.
The polarization method gives the highest precision. The other methods
are also important as they contribute to the determination of the
properties of the scalar top quark. For example, the scalar character of the 
stops can be established from the threshold cross section scan.

A new study for small mass difference, thus small visible energy, shows
that a Linear Collider has a large potential to study the scalar top production
and decay, in particular in this experimentally very challenging scenario.

From detailed simulations together with estimated errors for measurements 
in the neutralino/chargino sector,
the expected cosmological dark matter relic density can be
computed. The precision at a Linear Collider will be similar to the 
current precision of WMAP.
The uncertainty in the dark matter prediction from a Linear Collider is 
dominated by the precision of the scalar top quark mass measurement.

\vspace*{-1mm}
\begin{acknowledgments}
The authors are grateful to P.~Bechtle, S.~Mrenna and T.~Kuhl for practical advice.
\end{acknowledgments}

\end{document}